\documentclass[twocolumn]{aastex62}

\usepackage{graphics,graphicx,xspace,natbib,amssymb}
\usepackage[caption=false]{subfig}
\usepackage{amsmath}
\usepackage{threeparttable}
\usepackage{makecell}

\usepackage{multirow}
\usepackage{fontawesome5}

\newcommand{\squiggle}{SQuIGG$\vec{L}$E\xspace}
\newcommand{\oii}{[O\,{\sc ii}]\xspace}
\newcommand{\oiii}{[O\,{\sc iii}]\xspace}
\newcommand{\fburst}{$f_{\rm{burst}}$\xspace}
\newcommand{\tburst}{$t_{\rm{burst}}$\xspace}
\newcommand{\tq}{$t_q$\xspace}
\newcommand{\comment}[1]{}
\newcommand{\logsfrratio}{$\log{\rm{SFR}}_{\rm{ratio}}$\xspace}

\shorttitle{Recovering the star formation histories of recently-quenched galaxies}
\shortauthors{Suess et al.}

\begin{document}

\title{Recovering the star formation histories of recently-quenched galaxies:\\the impact of model and prior choices}

\author{Katherine A. Suess}
\affiliation{Department of Astronomy and Astrophysics, University of California, Santa Cruz, 1156 High Street, Santa Cruz, CA 95064 USA}
\affiliation{Kavli Institute for Particle Astrophysics and Cosmology and Department of Physics, Stanford University, Stanford, CA 94305, USA}

\author{Joel Leja}
\affiliation{Department of Astronomy \& Astrophysics, The Pennsylvania State University, University Park, PA 16802, USA}
\affiliation{Institute for Computational \& Data Sciences, The Pennsylvania State University, University Park, PA, USA}
\affiliation{Institute for Gravitation and the Cosmos, The Pennsylvania State University, University Park, PA 16802, USA}

\author{Benjamin D. Johnson}
\affiliation{Center for Astrophysics | Harvard \& Smithsonian, 60 Garden Street, Cambridge, MA 02138, USA}

\author{Rachel Bezanson}
\affiliation{Department of Physics and Astronomy and PITT PACC, University of Pittsburgh, Pittsburgh, PA, 15260, USA} 

\author{Jenny E. Greene}
\affiliation{Department of Astrophysical Sciences, Princeton University, Princeton, NJ 08544, USA}

\author{Mariska Kriek} 
\affiliation{Leiden Observatory, Leiden University, P.O.Box 9513, NL-2300 AA Leiden, The Netherlands}
\affiliation{Astronomy Department, University of California, Berkeley, CA 94720, USA}

\author{Sidney Lower}
\affiliation{Department of Astronomy, University of Florida, 211 Bryant Space Science Center, Gainesville, FL, 32611, USA}

\author{Desika Narayanan}
\affiliation{Department of Astronomy, University of Florida, 211 Bryant Space Science Center, Gainesville, FL, 32611, USA}
\affiliation{University of Florida Informatics Institute, 432 Newell Drive, CISE Bldg E251 Gainesville, FL, 32611, US}
\affiliation{Cosmic Dawn Centre at the Niels Bohr Institue, University of Copenhagen and DTU-Space, Technical University of Denmark}

\author[0000-0003-4075-7393]{David J. Setton}
\affiliation{Department of Physics and Astronomy and PITT PACC, University of Pittsburgh, Pittsburgh, PA, 15260, USA} 

\author[0000-0003-3256-5615]{Justin~S.~Spilker}
\affiliation{Department of Physics and Astronomy and George P. and Cynthia Woods Mitchell Institute for Fundamental Physics and Astronomy, Texas A\&M University, 4242 TAMU, College Station, TX 77843-4242, US}

\email{suess@ucsc.edu}

\begin{abstract}
Accurate models of the star formation histories (SFHs) of recently-quenched galaxies can provide constraints on when and how galaxies shut down their star formation. The recent development of ``non-parametric" SFH models promises the flexibility required to make these measurements. However, model and prior choices significantly affect derived SFHs, particularly for post-starburst galaxies (PSBs) which have sharp changes in their recent SFH. In this paper, we create mock PSBs, then use the \texttt{Prospector} SED fitting software to test how well four different SFH models recover key properties. We find that a two-component parametric model performs well for our simple mock galaxies, but is sensitive to model mismatches. The fixed- and flexible-bin non-parametric models included in \texttt{Prospector} are able to rapidly quench a major burst of star formation, but systematically underestimate the post-burst age by up to 200~Myr. We develop a custom SFH model that allows for additional flexibility in the recent SFH. Our flexible non-parametric model is able to constrain post-burst ages with no significant offset and just $\sim90$~Myr of scatter. Our results suggest that while standard non-parametric models are able to recover first-order quantities of the SFH (mass, SFR, average age), accurately recovering higher-order quantities (burst fraction, quenching time) requires careful consideration of model flexibility. These mock recovery tests are a critical part of future SFH studies. Finally, we show that our {new, public} SFH model is able to accurately recover the properties of mock star-forming and quiescent galaxies and is suitable for broader use in the SED fitting community. \href{https://github.com/bd-j/prospector}{\faGithubSquare}
\end{abstract}


\section{Introduction}

One of the largest unsolved problems in galaxy evolution is understanding the buildup of quiescent galaxies over cosmic time: when and why do galaxies ``quench" and cease forming stars? Understanding the star formation histories (SFHs) of quiescent galaxies is a critical piece of this puzzle. Robust SFHs constrain two properties: how long a galaxy has been quenched, and how long it took for the galaxy to transition from star-forming to quiescent. 

Different proposed quenching mechanisms operate on different timescales--- for example, major mergers and black hole feedback could quench galaxies on relatively short timescales, whereas mechanisms that rely on reducing halo accretion rates operate over longer timescales \citep[e.g.,][]{keres05,feldmann15,wright19,rodriguez-montero19}. Measuring how rapidly star formation ceased can thus help constrain what mechanisms were responsible for the shutdown. Quantifying how long galaxies have been quenched allows us to construct a timeline of how various galaxy properties evolve after star formation shuts down. Because it is impossible to watch a single galaxy evolve through the quenching process, cross-sectional studies using accurate post-quenching ages are the only way to gain an understanding of how galaxy structure, AGN activity, molecular gas contents, and other key properties change throughout the quenching process \citep[e.g.,][]{french18,bezanson22}. 

Major classes of recently-quenched galaxies include ``green valley" galaxies, which appear to quench gradually \citep[e.g.,][]{martin07,mendez11,schawinski14,wu18}, and ``post-starburst" galaxies (PSBs), which are thought to quench rapidly after a major burst of star formation \citep[for a recent review, see][]{french21}. In this work, we concentrate on accurately measuring the SFHs of PSBs. The unique B5V/A-star dominated spectra of these galaxies make them relatively easy to identify in both photometric and spectroscopic surveys. While they are present across redshift, PSBs represent the dominant formation pathway for quiescent galaxies above $z\sim1-2$ \citep[e.g.,][]{whitaker12_psb,wild16,rowlands18,belli19}. Because these galaxies are thought to quench after a major burst of star formation, SFH models for PSBs must be able to (a) capture early star formation before the recent burst, (b) produce a large recent burst of star formation with variable duration and burst mass fraction, and (c) rapidly shut down the burst while constraining the time since quenching. This rapid evolution and large SFR dynamic range mean that, in many ways, PSBs represent one of the most difficult test cases for SFH models. Models that are able to describe the extreme SFHs of PSBs are likely to have sufficient flexibility to describe the vast majority of galaxy SFHs across redshift. 

Accurately measuring the SFHs of PSBs from multi-wavelength data is challenging. 
Historically, most spectral energy distribution (SED) fitting codes have assumed a relatively simple parametric form for the SFH that depends on a small handful of parameters \citep[for a review, see][]{walcher11,conroy13}. These parametric forms impose strong priors on specific star formation rates (sSFRs) and mass-weighted ages, and therefore results from parametric SFH fits may not accurately reflect the true mass assembly histories of galaxies \citep[e.g.,][]{carnall19,lower20}. The most widely-used parametric model is the delayed-$\tau$ model, where SFR~$\propto te^{-t/\tau}$ and the timescale $\tau$ is a free parameter. This type of SFH model inextricably links the ongoing SFR, the recent SFR, and the SFR at very early times. This means that these parametric models have particular difficulties with the extreme SFHs of PSBs: they cannot easily reproduce both a strong recent starburst and low ongoing SFRs. Furthermore, standard parametric SFHs do not allow for both an old component and a recent burst in these galaxies: all of the mass is forced into the recent burst, likely an unphysical solution.

Several recent works have mitigated these difficulties by describing PSB SFHs as the sum of multiple parametric components. \citet{kaviraj07} allowed for both an old and young component by modeling PSB SFHs as the sum of an instantaneous burst at high redshift and an exponential recent burst. Similarly, \citet{french18} modeled PSB SFHs as an old linear exponential component in addition to either one or two recent exponential bursts and \citet{wild20} assumed that PSB SFHs can be described as the sum of an old  exponentially-declining component and a recent double-powerlaw burst. All three of these approaches allow for a varying fraction of the mass to be formed in the recent burst versus the underlying older component, solving one of the main issues with using delayed-$\tau$ models for PSBs. However, these approaches still explicitly assume a parametric form for both the older component and the burst.

Additional flexibility in the shapes of galaxy SFHs has recently been made possible through advances in inference techniques allowing higher dimensional models: these ``non-parametric" SFHs do not assume a specific analytic form for the SFH but instead allow for arbitrary SFRs in adjacent timebins (e.g., \citealt{conroy13,iyer17,iyer19,leja19a,leja19b}; see also \citealt{alarcon22} for a flexible physically-motivated parametric model). Non-parametric models introduce a larger number of free parameters into the fit in exchange for more freedom and flexibility in the derived SFHs. This additional freedom allows for non-parametric SFHs to more accurately reproduce the SFHs of simulated galaxies, leading to more accurate recovery of quantities such as stellar mass \citep{lower20}. Stellar mass functions derived from non-parametric SFH fitting are also more consistent with the observed star formation rate density of the universe \citep{leja20}. In theory, these non-parametric models provide great promise for accurately reproducing PSB SFHs.

However, even with non-parametric SFHs there are many possible ways to mathematically describe the SFH model and place priors on the fit variables. Just like parametric SFHs, these non-parametric model choices can have impacts on derived quantities such as stellar mass and SFR \citep[e.g.,][]{iyer17,lower20}. \citet{leja19a} tests how well different non-parametric priors are able to recover the properties of mock galaxies using the \texttt{Prospector} SED fitting code \citep{johnson20}. Notably, they find that the choice of prior is the {\it primary determinant} of the shape of the SFH posterior, more impactful than even the photometric noise \citep{leja19a}. Furthermore, the total number of additional free parameters that can be added to these fits is still limited: as the dimensionality of the fit increases, so does the computational time. As more studies begin to use these new non-parametric SFH fitting tools to constrain the quenching times of galaxies \citep[e.g.,][]{estrada-carpenter20,tacchella21,belli21,akhshik21,werle22}, the need for a detailed study of the effects of non-parametric priors on the SFHs of recently-quenched galaxies is clear.

In this paper, we test how well different SFH models are able to recover the properties of mock PSBs. Our mock PSBs are created with an SFH consisting of an older delayed-$\tau$ component plus a recent tophat burst. These relatively simple inputs allow us to understand the impact of different SFH model and prior choices on output quantities of interest, including stellar mass, ongoing star formation rate (SFR), burst mass fraction, and quenching time. Our goal is to understand biases in these recovered quantities and identify the best model for recovering the SFHs of recently-quenched galaxies. We test three different non-parametric SFH models: two ``out-of-the box" non-parametric models included in the public \texttt{Prospector} distribution, and one non-parametric SFH specifically designed for PSBs {(now part of the public \texttt{Prospector} distribution \href{https://github.com/bd-j/prospector}{\faGithubSquare}).} We also test a pararametric SFH model consisting of two delayed-$\tau$ components, similar to the models used in previous PSB SFH studies. This double delayed-$\tau$ model is nearly identical to the SFH used to create our mock PSBs, and allows us to investigate how well parametric SFH models fare in a ``best-case" scenario where the model assumptions match the true SFHs. 

Section~\ref{sec:mock_generation} describes our mock PSB data; while here we focus on SDSS-quality spectra and photometry such as those available for the \squiggle PSB survey \citep{suess20}, these mocks are similar to the data that can be expected from upcoming spectroscopic surveys such as DESI, PFS, and MOONRISE. In Section~\ref{sec:model_priors}, we describe our SED fitting setup and our four SFH models in detail. Section~\ref{sec:pick_model} determines the best SFH model to use for PSBs, and Section~\ref{sec:sfqui} shows that this model is also able to accurately reproduce the SFHs of quiescent and star-forming galaxies. 

Throughout this paper we assume a flat $\Lambda$CDM cosmology with $\Omega_{\rm m}=0.3$, $\Omega_\Lambda=0.7$, and $h=0.7$. Stellar masses are quoted assuming a \citet{chabrier03} initial mass function. For consistency with other SED fitting works, stellar masses $\log({M_*}/M_\odot$) are quoted in units of the surviving stellar mass-- e.g., accounting for mass loss; all moments of the SFH including the ongoing SFR and the burst mass fraction are quoted in units of total mass formed.

\section{Generating mock spectra}
We generate two sets of mock data for this paper. In Section~\ref{sec:mock_generation}, we describe the generation of mock PSB spectra. These are used in Section~\ref{sec:pick_model} to identify the best model to recover PSB SFHs. Section~\ref{sec:mock_sfqui} describes the generation of mock quiescent and star-forming spectra; these are used in Section~\ref{sec:sfqui} to verify that the PSB SFH model is suitable for broader use.

\subsection{Mock PSB spectra}
\label{sec:mock_generation}

We create a large grid of mock SDSS-like optical spectroscopy and photometry using FSPS \citep{conroy09,conroy10}. All mock galaxies assume a \citet{chabrier03} IMF, the \citet{calzetti00} dust law, a total formed stellar mass of $10^{11.25}M_\odot$, and a velocity dispersion of 200 km/s. After taking mass loss into account, this total mass formed equates to a surviving stellar mass of $10^{11.05-11.10}$ depending on metallicity and SFH. We include nebular emission in all mock spectra using the default FSPS parameters.

We vary the dust attenuation values, stellar metallicites, star formation histories, and spectral S/N of the mocks. Dust extinction varies between zero and 1.5 magnitudes. Following, e.g., \citet{wild20}, we double the dust attenuation around young stars. Metallicity varies between solar and 0.5~dex above solar \citep[as expected for massive galaxies, e.g.][]{gallazzi05}. We model the SFHs of the mock galaxies with two components: an older delayed-$\tau$ model plus a recent tophat burst. We vary the mass fraction in the recent burst (\fburst), the duration of the recent burst (\tburst), the time since quenching (\tq), and the star formation rate after quenching (e.g., the amount of ``frosting", SFR$_q$). 

We choose the 10th, 33rd, 66th, and 90th percentile noisiest galaxies in the \squiggle sample \citep{suess22} to use as noise templates. We will use the error spectrum, redshift, and wavelength coverage of these noise templates as guides to ensure that the properties of our mock spectra are a good match to observed data quality. 

Table \ref{table:recovery_values} shows the values of each parameter that we vary to create our grid of mock SDSS-quality spectra. Generating a mock spectrum for every grid location would be immensely time-consuming--- this would produce $\sim$60,000 mock spectra. We therefore randomly select 5,000 points on the grid to generate mock spectra. 

\begin{table}[]
\caption{Values used to generate the grid of mock PSB spectra. Values are chosen to roughly span the range probed by the \squiggle sample of intermediate-redshift PSBs \citep{suess22}.}
\label{table:recovery_values}
\begin{tabular}{|l|p{.5\linewidth}|} \hline
parameter            & values                                                                                                           \\ \hline \hline
\fburst  & 0.1, 0.2, 0.5, 0.7, 0.9, 0.99                                                                                    \\ \hline
\tq & 0.05, 0.1, 0.2  0.3, 0.4, 0.5, 0.6, 0.8, 1.0 Gyr                      \\ \hline
metallicity          & $\log{Z/Z_\odot}$ = 0.0, 0.2, 0.5                                                                                           \\ \hline
dust A$_v$           & 0.0, 0.5, 1.0, 1.5 mag                                                                                           \\ \hline
SFR$_q$  & 1e-5, 1e-3, 1e-2, 1e-1, 2, 4, 6, 8, 10, 20, 30 $M_\odot \rm{yr}^{-1}$ \\ \hline
burst duration       & 100, 200, 400, 600 Myr                                                                                           \\ \hline
spectral S/N                  & 6.2, 7.0, 7.8, 9.7    \\    \hline
\end{tabular}
\end{table}

We redshift each mock spectrum to the same redshift as its noise template, broaden the spectral resolution to match the wavelength-dependent instrumental dispersion of the template SDSS spectrum, then interpolate the mock FSPS spectra onto the same wavelength grid as the template. 
Next, we 
add random Gaussian noise to the mock spectrum following the per-pixel S/N of the template SDSS spectrum. 
We also generate mock photometry for each galaxy in the SDSS and WISE bands. We perturb the mock photometry, again with a random Gaussian scaled by the true S/N of the SDSS and WISE observations of the template spectrum.

After this process, we have a total of 5,000 mock galaxies with SDSS-quality data. We then run the \squiggle color-based PSB selection method on these mock spectra. 1,821/5,000 of these mock galaxies meet the \squiggle PSB selection criteria. \citet{suess22} explores in more detail the types of mock galaxies that satisfy the \squiggle sample criteria; these PSB-like mock galaxies tend to have low ongoing SFRs, relatively little dust obscuration, and a range of burst fractions and quenching timescales.

\subsection{Mock star-forming and quiescent spectra}
\label{sec:mock_sfqui}
In Section~\ref{sec:sfqui}, we will use mock star-forming and quiescent spectra to ensure that the SFH model we develop and test for PSBs is suitable for broader use. Our main goal is to verify that the PSB SFH model is able to reproduce a broad range of ongoing SFRs and does not artificially create recent starbursts in galaxies that did not experience them. Therefore, we create relatively simple mock star-forming and quiescent galaxies based off of the best-fit FAST \citep{kriek09} SED fitting parameters of observed galaxies from the 3D-HST survey \citep{skelton14,momcheva16}. This allows us to select stellar masses, ongoing SFRs, and dust attenuation values that are realistic for a population of massive intermediate-redshift galaxies. More detailed testing of a wide range of SFHs would likely require mock observations of simulated galaxies \citep[e.g.,][]{lower20}, which is beyond the scope of this paper.  

We select all galaxies in the 3D-HST master catalog \citep{skelton14,momcheva16} with a best-fit redshift $0.5\le z_{\rm{best}}\le 1.0$, a best-fit stellar mass $\log{M_*/M_\odot}\ge10.75$, and a `use\_phot' flag equal to one. These mass and redshift limits are similar to those of the \squiggle survey that we base our mock PSB galaxies on (Section~\ref{sec:mock_generation}). From these 487 galaxies, we randomly select 100 to serve as templates for our mock star-forming and quiescent galaxies. Roughly half of these 100 galaxies are identified as quiescent from their $UVJ$ colors, while the other half are actively star-forming. We again use FSPS to create mock spectra using the best-fit redshift, $A_v$, and $\log{M_*/M_\odot}$ of each galaxy. The star formation history is modeled using a delayed-$\tau$ function using the best-fit $\tau$ and age from the 3D-HST FAST fit. We assume solar metallicity and a fixed \citet{calzetti00} attenuation curve (corresponding to a \citealt{kriek13} dust index of zero) for all the mocks, as these were the parameters assumed in the 3D-HST FAST fits. We assume a velocity dispersion of 200~km/s and broaden the spectra according to the SDSS instrumental dispersion.
After creating the mock spectrum, we add realistic noise following the same procedure used for the PSB galaxies. We pick the \squiggle spectrum at the closest redshift to the 3D-HST mock, then perturb the spectrum and photometry within the observed error bars of the \squiggle spectrum.

\section{SED fitting model and priors}
\label{sec:model_priors}

\renewcommand{\arraystretch}{1.2}
\begin{table*}[]
\caption{Description of parameters and priors used common to all \texttt{Prospector} fits.}
\label{table:priors}
\begin{tabular}{|l|l|p{.3\linewidth}|p{.5\linewidth}|} \hline
            & Parameter                     & Description                                                                                            & Prior                                                                                                                                                           \\ \hline \hline
           \multirow{4}{*}{} & $\log\frac{\rm{M}_*}{\rm{M}_\odot}$ & total stellar mass formed                                                                              & uniform: min = 9.5, max = 12.5                                                                                                                                  \\ \cline{2-4}
            & $\log\frac{\rm{Z}_*}{\rm{Z}_\odot}$   & stellar metallicity                                                                                    & clipped normal: min = -0.5, max = 1.0,\newline mean and $\sigma$ following Leja+19b mass-metallicity prior                   \\ \cline{2-4}
            & $\sigma$                      & stellar velocity dispersion                                                                            & uniform: min = 100, max = 300 km/s                                                                                                                   \\ \cline{2-4}
            & $z$                           & redshift                                                                                               & {\bf fixed} to SDSS spectroscopic redshift                                                                                                     \\ \hline \hline
\multirow{6}{*}{dust}  & $\hat\tau_{\lambda,2}$        & diffuse dust optical depth                                                                             & uniform: min 0.0 mag, max 2.5 mag                                                                                                                               \\ \cline{2-4}
            & $\hat\tau_{\lambda,1}$        & birth-cloud dust optical depth                                                                         & {\bf fixed} to  $\hat\tau_{\lambda,2}$\newline (e.g., young stars are attenuated twice as much as old stars) \\ \cline{2-4}
            & $n$                           & slope of Kriek \& Conroy dust law                                                                      & uniform: min -1.0, max 0.4                                                                                                                                      \\ \cline{2-4}
            & $\gamma_e$                   & warm dust fraction                                                                                     & {\bf fixed} to 0.01                                                                                                                            \\ \cline{2-4}
            & U$_{\rm min}$                 & minimum radiation field to which dust is exposed            & {\bf fixed} to 1.0                                                                                                                             \\ \cline{2-4}
            & q$_{\rm PAH}$                  & PAH mass fraction                                                                                      & {\bf fixed} to 2.0\%                                                                                                                              \\ \hline \hline
\multirow{3}{*}{noise} & $j_{\rm{spec}}$               & spectroscopic jitter term                                                                              & uniform: min = 1.0, max = 1.5                                                                                                                                   \\ \cline{2-4}
            & $f_{\rm{out}}$                & fraction of pixels in spectrum considered to be outliers    & uniform: min = 0, max = 0.5                                                       \\ \cline{2-4}
            & $s_{\rm{out}}$                & increased noise for spectral outliers                                                                  & {\bf fixed} to 5.0           \\\hline                                                                                                                 
\end{tabular}
\end{table*}

We use the \texttt{Prospector} stellar population synthesis code \citep{johnson17, leja17, johnson20} to simultaneously fit the SDSS spectra and the SDSS and WISE photometry of all galaxies in our mock samples. Our general setup is the same as in \citet{suess22}. Table~\ref{table:priors} lists the free parameters and priors that are used for all SFH models tested; Section~\ref{sec:sfh_priors} and Table~\ref{table:sfh_priors} describe the free parameters and priors used for each of the four SFH models we test. 

We use the Flexible Stellar Population Synthesis \citep[FSPS;][]{conroy09,conroy10} library to generate stellar populations, and the  \texttt{dynesty} dynamic nested sampling package \citep{speagle20} to sample posteriors. We adopt the MILES spectral library \citep{miles} and the MIST isochrones \citep{dotter16,choi16}; the MIST isochrones are generated with MESA \citep{paxton11,paxton13,paxton15,paxton18}. We assume the \citet{chabrier03} initial mass function, fix the model redshift to the spectroscopic redshift of the mock galaxy, and add nebular emission to the spectra using the default fixed parameters in \texttt{Prospector}. 

Total stellar mass formed, metallicity, and velocity dispersion are free in our fits. We allow $\log{\rm{M}_*/\rm{M}_\odot}$ to vary between $9.5$ and $12.5$. We adopt the mass-metalliticy prior described in \citet{leja19b}, where the $\log{Z/Z_\odot}$ prior is a clipped normal distribution with a minimum of $-0.5$ and a maximum of $1.0$. The mean and $\sigma$ of the prior is set based on the total stellar mass, following a modified version of the \citet{gallazzi05} local mass-metallicity relation. We fit for the velocity dispersion using a flat prior between $100-300$~km/s.

Additionally, we fit for several parameters designed to prevent inaccurate calibration or bad pixels from skewing the output. As described in \citet{johnson20}, we include a free spectroscopic jitter term with a uniform prior between 1.0 and 1.5; this multiplicative term increases the noise in the spectrum. We also use \texttt{Prospector}'s pixel outlier model, which allows for a fraction $f_{\rm{outlier}}$ of pixels to have their uncertainties underestimated by a factor of $s_{\rm{outlier}}$. $f_{\rm{outlier}}$ is free, with a uniform prior between $10^{-5}$ and 0.5; $s_{\rm{outlier}}$ is fixed to 5.0. Finally, we use the polynomial SED model in \texttt{Prospector}, which optimizes out a low-order polynomial with every likelihood call; this is intended to account for any calibration issues with the spectra, and effectively upweights the lines as compared to the shape of the spectral continuum. 

We mask all spectral pixels within $50\AA$ of the 3727$\AA$ \oii line or within $100\AA$ of the $5007\AA$ \oiii line. In real post-starburst galaxies, these lines are often contaminated by LINER or AGN emission \citep[e.g.,][]{lemaux10,yan06,greene20}. While our mock galaxies do not include this non-stellar emission, we want the tests in this paper to be as relevant as possible for fitting observed post-starburst galaxies such as those in \citet{suess22}. Therefore, we exactly replicate the emission line masking performed in that work. 

We use the \citet{kriek09} dust law with a free slope and optical depth. We place a uniform prior on the dust law slope between -1 and 0.4, and a uniform prior on the diffuse dust optical depth between 0.0 and 2.5 magnitudes. Following \citet{wild20}, we fix the birth-cloud optical depth to the same value as the diffuse optical depth. This implies that young stars are attenuated twice as much as old stars. Following \citet{leja19a}, we also set the dust emission parameters such that the warm dust fraction is fixed to 0.01, the minimum radiation field is fixed to 1.0, and the PAH mass fraction is fixed to 2\%. 

\subsection{SFH model}
\label{sec:sfh_priors}

\begin{figure*}
    \centering
    \includegraphics[width=\textwidth]{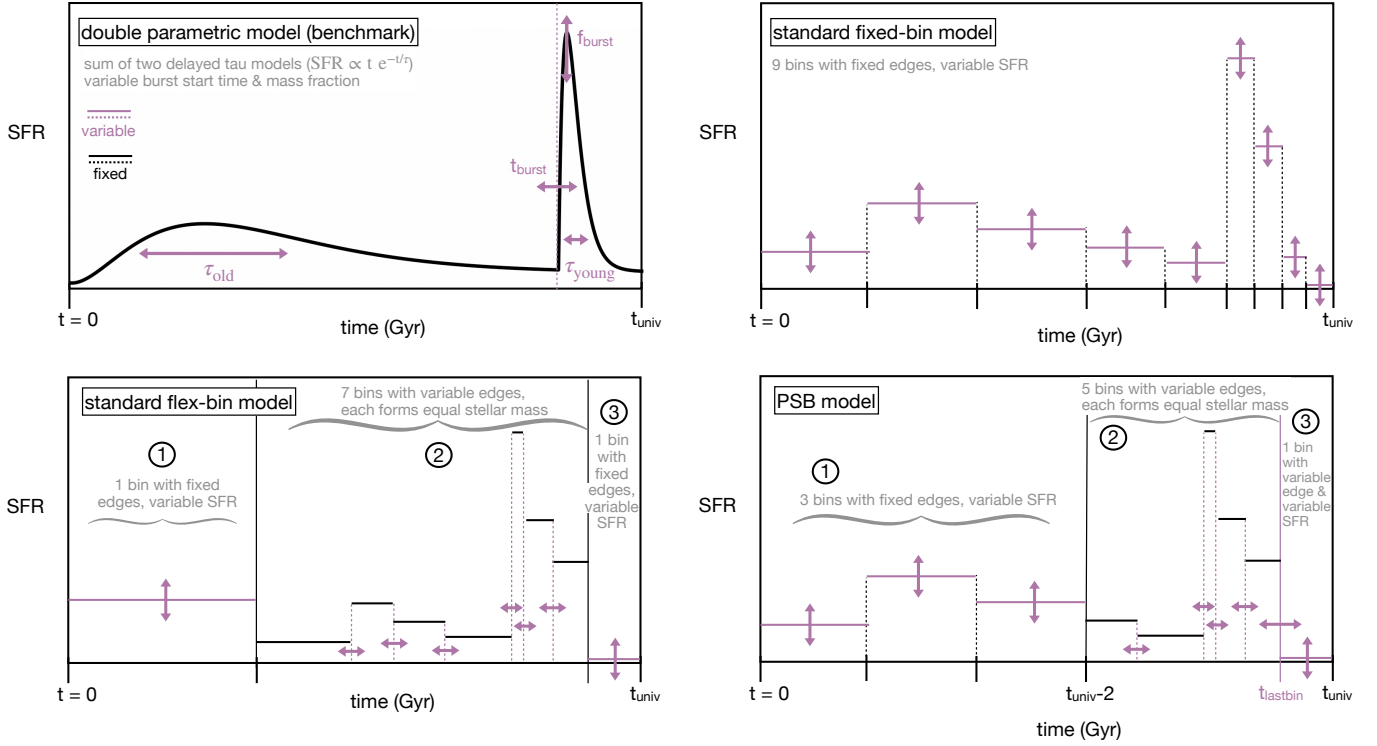}
    \caption{Illustrations of the four different star formation histories explored in this paper. The top left shows a parametric star formation history consisting of the sum of two delayed-$\tau$ models with different star formation timescales and a variable burst fraction. This double parametric model is similar to those used by \citet{french18} and \citet{wild20} to model PSB SFHs, though the exact parameterization differs. {Because this model is very similar to the SFH used to generate our mock spectra, it is expected to perform well by construction; this benchmark allows us to understand the relative effects of mock data quality and SFH parameterization.} The upper right and lower left panels show two of the ``standard" \texttt{Prospector} non-parametric star formation histories included as template libraries in the code \citep[see][]{leja19a}. The upper right shows the fixed-bin model, where the bin edges are fixed and the SFRs are allowed to vary; the lower left shows the flexible-bin model, where the bin edges are allowed to vary such that each bin forms equal stellar mass. In the flexible-bin model, the first and last bins remained fixed in order to allow for low instantaneous sSFRs \citep[see][]{leja19a}. The lower right panel shows the model that we specifically design for PSBs: it consists of three fixed-edge bins, five flexible bins, and one final bin with variable length and SFR that is intended to capture post-quenching star formation.}
    \label{fig:sfh_models}
\end{figure*}

In this work, we test four different SFH models. The first is a parametric model: the entire SFH is specified by a small handful of physical parameters. This model assumes that the galaxy SFH follows a specific functional form. The remaining three models are non-parametric models. These models typically assume that the SFR is a piecewise function and fit for the SFR in each adjacent timebin. While these models have significantly more flexibility (and more free parameters) than traditional parametric models, they do still require choices about priors; as for parametric models, these choices affect the output SFHs \citep[e.g.,][]{leja19a,lower20}. All four models are illustrated graphically in Figure~\ref{fig:sfh_models}, and a table of the parameters and priors used in each of the four models is shown in Table~\ref{table:sfh_priors}. Throughout the rest of this paper, we will explore the impact that these SFH model choices have on the output parameters of the fit.

\begin{table*}[]
\caption{Description of the parameters and priors used in each of the four SFH models tested in this paper. All Student $t$ priors are centered at the UniverseMachine expectations for a quiescent galaxy of similar mass and redshift, as described in the text. All SFH models are normalized using the total stellar mass, which is also a free parameter in our \texttt{Prospector} fits.}
\label{table:sfh_priors}
\begin{tabular}{|l|l|l|l|}
\hline
SFH model                               & $N_{\rm{params}}$  & Parameter Name                                                                    & Prior                             \\ \hline \hline
\multirow{4}{*}{double parametric}      & \multirow{4}{*}{4} & $t_{\rm{burst}}$: time when young component begins {[}Gyr{]}                      & uniform {[}0,  $t_{\rm{univ}}${]} \\ \cline{3-4} 
                                        &                    & $\tau_{\rm{old}}$: old component SF timescale {[}Gyr$^{-1}${]}                    & log-uniform {[}0.01,  30{]}        \\ \cline{3-4} 
                                        &                    & $\tau_{\rm{young}}$: young component SF timescale {[}Gyr$^{-1}${]}                & log-uniform {[}0.01,  30{]}        \\ \cline{3-4} 
                                        &                    & $f_{\rm{burst}}$: fraction of mass formed in young component                      & uniform {[}0,  1{]}               \\ \hline
standard fixed-bin                      & 8                  & log(SFR$_{\rm{ratio}})$: 8-vector; ratio of SFR in adjacent bins                  & Student $t$                       \\ \hline
\multirow{3}{*}{standard flex-bin}      & \multirow{3}{*}{8} & log(SFR$_{\rm{ratio,young}})$: ratio of SFR in youngest bin to last flex bin      & Student $t$                       \\ \cline{3-4} 
                                        &                    & log(SFR$_{\rm{ratio,old}})$: ratio of SFR in old bin to first flex bin            & Student $t$                       \\ \cline{3-4} 
                                        &                    & log(SFR$_{\rm{ratio}})$: 6-vector; ratio of SFR in flex bins                      & Student $t$                       \\ \hline
\multirow{4}{*}{PSB}                    & \multirow{4}{*}{9} & log(SFR$_{\rm{ratio,young}})$: ratio of SFR in youngest bin to last flex bin      & Student $t$                       \\ \cline{3-4} 
                                        &                    & log(SFR$_{\rm{ratio,old}})$: 3-vector; ratio of SFR in old bins to first flex bin & Student $t$                       \\ \cline{3-4} 
                                        &                    & log(SFR$_{\rm{ratio}})$: 4-vector; ratio of SFR in flex bins                      & Student $t$                       \\ \cline{3-4} 
                                        &                    & $t_{\rm{last}}$: width of last timebin {[}Gyr{]}                                  & uniform {[}0.01,  1.0{]}          \\ \hline
\end{tabular}
\end{table*}

\subsubsection{Parametric model: double delayed-$\tau$}
The first SFH that we test is a parametric model. Parametric models--- typically, single-component delayed-$\tau$ models--- are one of the most commonly-used SFHs when creating catalogs of stellar population properties for large samples of galaxies. However, a single-component parametric model is clearly unsuitable for recovering the SFHs of our mock PSBs: with only a single parametric component, all of the mass is forced into the recent burst and the model is unable to produce a range of burst mass fractions. Instead, similar to previous PSB SFH studies \citep[e.g.,][]{kaviraj07,french18,wild20} we use an SFH model that consists of the sum of two delayed-$\tau$ models. This SFH includes both an old and a young component, which allows for a variable fraction of the galaxy's total stellar mass to have been formed in the recent starburst. 

The old component is described by:
\begin{equation}
    \mathrm{SFR}_{\mathrm{old}}(t) \propto\ t e^{-t/\tau_{\rm{old}}},
\end{equation}
and the young component is described by:
\begin{equation*}\label{eqn:youngdel}
\quad \mathrm{SFR}_{\rm{y}}(t)\ \propto\ \begin{cases}
    \ \ \  (t - t_{\rm{burst}})\ e^{- (t - t_{\rm{burst}})/\tau_{\rm{young}}} & \quad  t > t_{\rm{burst}}\\
    \ \ \ 0 & \quad  t < t_{\rm{burst}}.
\end{cases}
\end{equation*}

The total SFH is the sum of these two components, weighted by the burst mass fraction:
\begin{equation}
    \mathrm{SFR}(t) \propto (1-f_{\rm{burst}}) \times \mathrm{SFR}_{\rm{old}} + f_{\rm{burst}} \times \mathrm{SFR}_{\rm{y}}.
\end{equation}

We place uniform priors on both $\tau_{\rm{young}}$ and $\tau_{\rm{old}}$ between 0.01 and 30 Gyr$^{-1}$, and allow $t_{\rm{burst}}$ to vary between 0~Gyr and the age of the universe. We place a uniform prior on the burst mass fraction between 0 and 1. 
We note that this SFH model is very similar to the model we use to generate our mock galaxies in Section~\ref{sec:mock_generation}; we therefore expect this model to recover the properties of the mock galaxies nearly perfectly.  

\subsubsection{Non-parametric model: fixed time bins}
The first of the three non-parametric models we test is the fixed-bin model preferred by \citet{leja19a}. In this model, the SFH is described by a piecewise function where the SFR is a constant in each of $N$ timebins. The edges of each timebin are fixed. The SFR in each fixed timebin is determined using the ``continuity" prior, which places a Student-t prior on the log of the ratio of the SFR in adjacent bins (``\logsfrratio"). This prior encourages smooth SFHs, where the SFR does not jump significantly between each timebin. However, sharp burst or quenching events are still allowed: the Student-t distribution has significantly more weight in the wings than a gaussian prior, meaning that sharp SFR transitions are not fully excluded from consideration.

\citet{leja19a} use a Student-t prior on \logsfrratio centered at zero (e.g., the maximum prior probability occurs when the galaxy has a constant SFR across all cosmic time). In this work, we place a physically-motivated prior on the SFH by using UniverseMachine. UniverseMachine is a Bayesian code that uses an abundance-matching approach to relate galaxy and halo assembly; it predicts a host of galaxy physical properties, including the SFH and stellar mass of galaxies across cosmic time \citep{behroozi19}. The UniverseMachine public data release includes the predicted SFHs for quiescent galaxies as a function of stellar mass and redshift. Depending on the spectroscopic redshift of the galaxy to be fit, we load in the UniverseMachine predicted SFH for $M_*=10^{11}M_\odot$ galaxies that are quiescent at that redshift. We calculate the \logsfrratio required for each non-parametric SFH model to reproduce that SFH. We then set a Student-t prior for \logsfrratio at these UniverseMachine values, with a width of 0.3~dex and a degree-of-freedom equal to one. 

The general shape of these UniverseMachine predictions is similar to the delayed-$\tau$ model shown in Figure~\ref{fig:sfh_models}: they are relatively smooth, with a bulk of star-formation at early times trending towards lower SFRs at the time of observation. As a result, this UniverseMachine prior is {\it more conservative} than a flat prior would be at intermediate redshifts: it effectively upweights the amount of mass that galaxies can form at early times. This early-formed mass will be largely invisible at the time of observation due to the well-known ``outshining" effect, where young stars are more luminous than older, redder stellar populations \citep[e.g.,][]{papovich01}. This results in relatively high estimates of the stellar mass, and conservatively low estimates on the fraction of the total stellar mass formed in the recent burst.

We choose to use $N=9$ timebins in our non-parametric SFH. This number is a balance between computational complexity--- adding more free parameters makes fitting more time-intensive--- and accurately constraining when each galaxy quenched. Because in this model the SFR can only change at the edge of each bin, we want to have a relatively large number of bins during the $\sim500$~Myr before observation; this allows for different galaxies to quench at different times. We use the following recent bins:

\begin{equation}
\begin{aligned}
    0\ \rm{Myr} < t_{lookback} < 20\ \rm{Myr} \\
    20\ \rm{Myr} < t_{lookback} < 50\ \rm{Myr} \\
    50\ \rm{Myr} < t_{lookback} < 100\ \rm{Myr} \\
    100\ \rm{Myr} < t_{lookback} < 200\ \rm{Myr} \\
    200\ \rm{Myr} < t_{lookback} < 500\ \rm{Myr}
\end{aligned}    
\end{equation}

\citet{ocvirk06} suggests that logarithmic time separations are appropriate for separating different stellar populations; therefore, we distribute the remaining four timebins log-normally between $500$~Myr and the age of the universe at each galaxy's redshift.

\subsubsection{Non-parametric model: flexible time bins}
The second non-parametric model we test is the flexible-bin model from \citet{leja19a}. Again, this model is described by a piecewise function where the SFR is constant in each of $N=9$ timebins. The edges of the first and last timebin are fixed; however, the edges of the other 7 timebins are allowed to vary such that each bin forms an equal stellar mass. Thus, as shown in Figure~\ref{fig:sfh_models}, periods of high SFR are captured by short timebins and periods of low SFR are captured by longer time bins. This model potentially allows for more flexibility in the duration, start time, and end time of the recent burst: the SFR can change at an arbitrary time, as opposed to only changing at the edges of fixed timebins.

Again, we use the continuity prior, where we place a Student-t prior on the \logsfrratio in adjacent bins. We choose the most recent fixed bin to be 100~Myr long to allow for a low instantaneous SFR. Because we are primarily interested in the recent SFH, we set the first timebin to cover the first 1.5~Gyr of the galaxy's history. The remaining 7 flexible bins have variable widths that are adjusted with each likelihood call.

\subsubsection{PSB model}
The final non-parametric model we test is optimized for PSBs, and was used in \citet{suess22} to fit the \squiggle PSB sample. 
Our goals are for this SFH model to be able to produce a recent burst of star formation with variable start time, duration, and peak SFR; rapidly quench this recent burst; provide a robust estimate of how long the galaxy has been quenched; and allow for a variable fraction of the galaxy's stellar mass to be formed prior to the recent burst. 
We achieve these goals using a combination of the fixed and flexible time bin approaches described above. 

We divide the SFH into three parts, as shown in Figure~\ref{fig:sfh_models}. The oldest portion of the SFH, from the beginning of the universe to 2~Gyr before observation, is divided into three bins with fixed edges and variable SFR. The second portion of the SFH is divided into five flexible bins: the edges of the bins can vary, and each bin forms an equal amount of stellar mass. 
Finally, the most recent portion of the SFH is modeled by a single bin with variable SFR {\it and} a variable length. The inclusion of the fixed early-time bins allows for a significant fraction of the galaxy's mass to be formed at early times. As discussed in \citet{leja19a}, these fixed-edge bins also allow for lower sSFRs in the following flexible bins. The flexible period covers the `burst' portion of the post-starburst SFH. The variable bin widths in this section allow the burst start time and width to be determined by the data. The final bin is intended to capture any low levels of star formation taking place after the burst ends. The variable width of this final bin allows for quenching to occur at an arbitrary time instead of a fixed set of bin edges. 

As with the two other non-parametric SFH models tested in this paper, we set a continuity prior on \logsfrratio. We center the Student-t prior around the UniverseMachine estimates for a quiescent galaxy of similar mass and redshift, and use a width of 0.3~dex and a degree of freedom equal to one.

We note that this SFH model is now included in the public \texttt{Prospector} distribution as ``continuity\_psb\_sfh" in the template library.

\section{Identifying the best SFH model for post-starburst galaxies}
\label{sec:pick_model}

\begin{figure*}
    \centering
    \includegraphics[width=\textwidth]{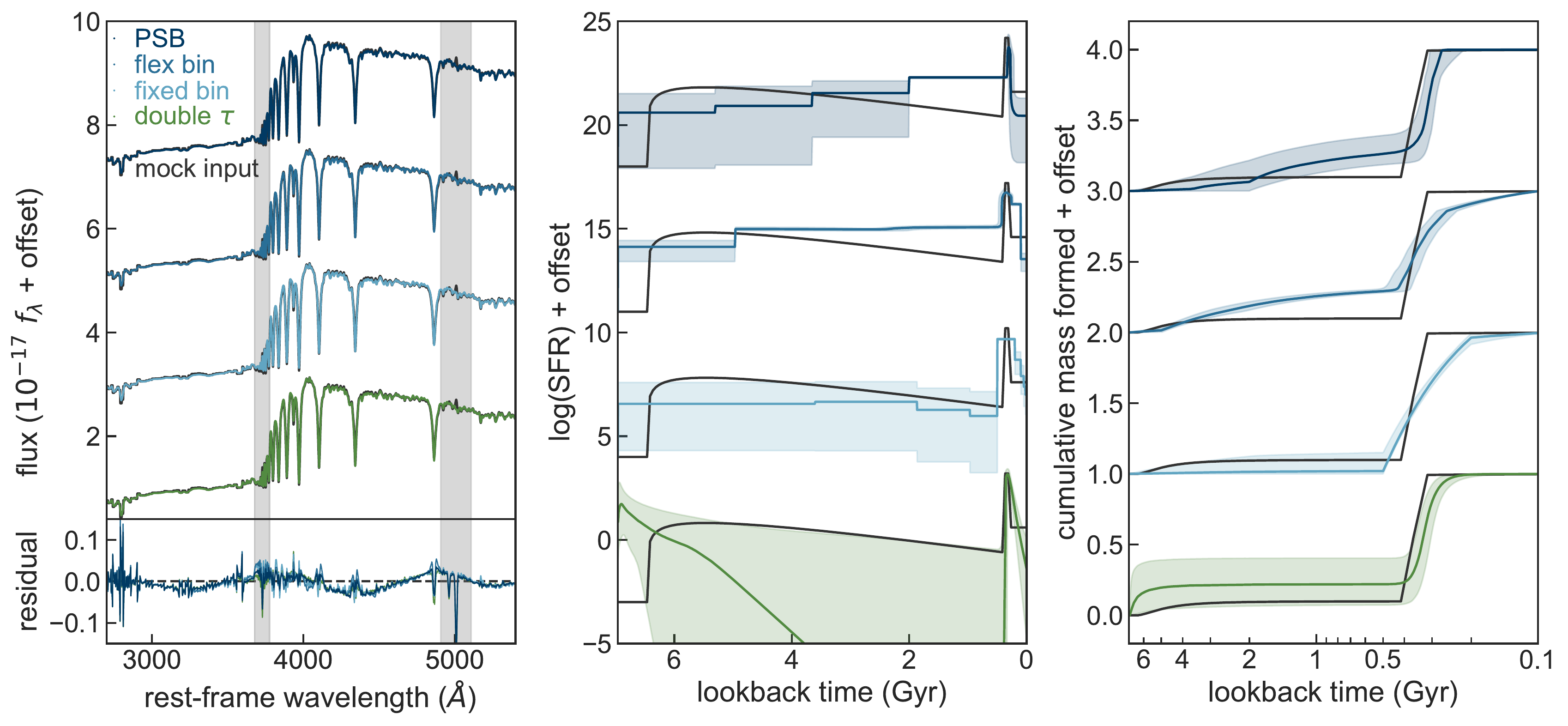}
    \caption{Example \texttt{Prospector} fits to one mock galaxy using each of the four SFH models we test; the input values are shown in grey, the median posterior values of the three non-parametric models are shown in shades of blue, and the median posterior values of the two parametric models are shown in shades of green. All values are shown with an arbitrary additive offset to improve visibility. The left panel shows the median posterior SDSS spectrum and the flux residuals. The shaded grey bars show the regions around the \oii and \oiii lines that are masked in the fits; the two parametric models show excess \oiii emission indicating overestimated ongoing SFRs. The middle panel shows the recovered SFH for each model. The two parametric models fail to capture either the shape of the early-time star formation or the rapid recent burst. The three non-parametric models recover early-time star formation well, but differ in how accurately they recover the shape and quenching time of the recent burst. The right panel shows the cumulative mass formation history for each SFH model (shown on a logarithmic lookback time scale to highlight the recent SFH). Both the input and recovered cumulative mass formed curves are normalized to form 100\% of the total stellar mass at the the time of observation.}
    \label{fig:exampleFit}
\end{figure*}

Here, we use the mock PSB spectra described in Section~\ref{sec:mock_generation} to test how well each SFH model described in Section~\ref{sec:sfh_priors} is able to recover various properties of interest, including stellar mass, dust attenuation, ongoing SFR, and time since quenching. We note that, critically, we generated mock spectra using the same dust law used in our \texttt{Prospector} fitting. This means that these mock recovery tests are {\it not} sensitive to any possible differences between our assumed dust model and the true dust law in observed PSBs. If true PSBs do not follow the \citet{kriek13} dust law that we assume here, then the systematic uncertainty in recovered properties could be larger than we find in these mock recovery tests. Testing which dust law best describes PSBs requires using real, not mock, observations, and is beyond the scope of this paper.

We randomly choose 300 mock PSB spectra and fit each of them with all four SFH models described in Section~\ref{sec:sfh_priors}. 
Figure~\ref{fig:exampleFit} shows an example of the fitting results for one mock PSB. The left panel shows the input spectrum and median posterior spectrum using each of four SFHs. For clarity of presentation, we add an offset to each spectrum so they do not overlap. All four models provide generally good agreement with the data. The center panel of Figure~\ref{fig:exampleFit} shows the input SFH (grey) and the recovered SFH (blue/green lines) for each SFH model. The shaded regions show the 16-84\% confidence interval around each SFH. The right panel of Figure~\ref{fig:exampleFit} shows the cumulative mass formation history for each model; this is simply an alternative view of the SFHs shown in the central panel. The center and right panels show that all three non-parametric models are able to capture some amount of early-time star formation as well as the steep recent burst. One of the major differences in the non-parametric models, explored further in Section~\ref{sec:burst_props}, is when they quench after the recent burst. 

In the rest of this section, we explore quantitatively how well each SFH model recovers the properties of all 300 mock PSBs that we fit.
We report all quantities as the median of the posterior distribution; 1$\sigma$ error bars are the 16th and 84th percentiles. For derived quantities such as time since quenching and mass-weighted age, we calculate the derived quantity for each posterior draw, then calculate the median and 1$\sigma$ error bar using the weights returned by the \texttt{dynesty} sampler. Median spectra and $M_{\rm formed}/M_{\rm surviving}$ are calculated using the 1,000 highest-weight posterior draws to save computational time; these draws contain the vast majority of the total posterior mass. 

We note that the output SFHs for both the flexible-bin model and the PSB model have different bin edges for each likelihood draw. For this reason, we interpolate each SFH draw onto a uniform 100~Myr spacing time grid before taking the weighted median and 16-84th percentile range. This interpolation causes the flexible-bin and PSB models to appear to have much higher time resolution than the fixed-bin model in Figure~\ref{fig:exampleFit}.

\subsection{Basic properties}

\begin{figure}[ht]
    \centering
    \includegraphics[width=.45\textwidth]{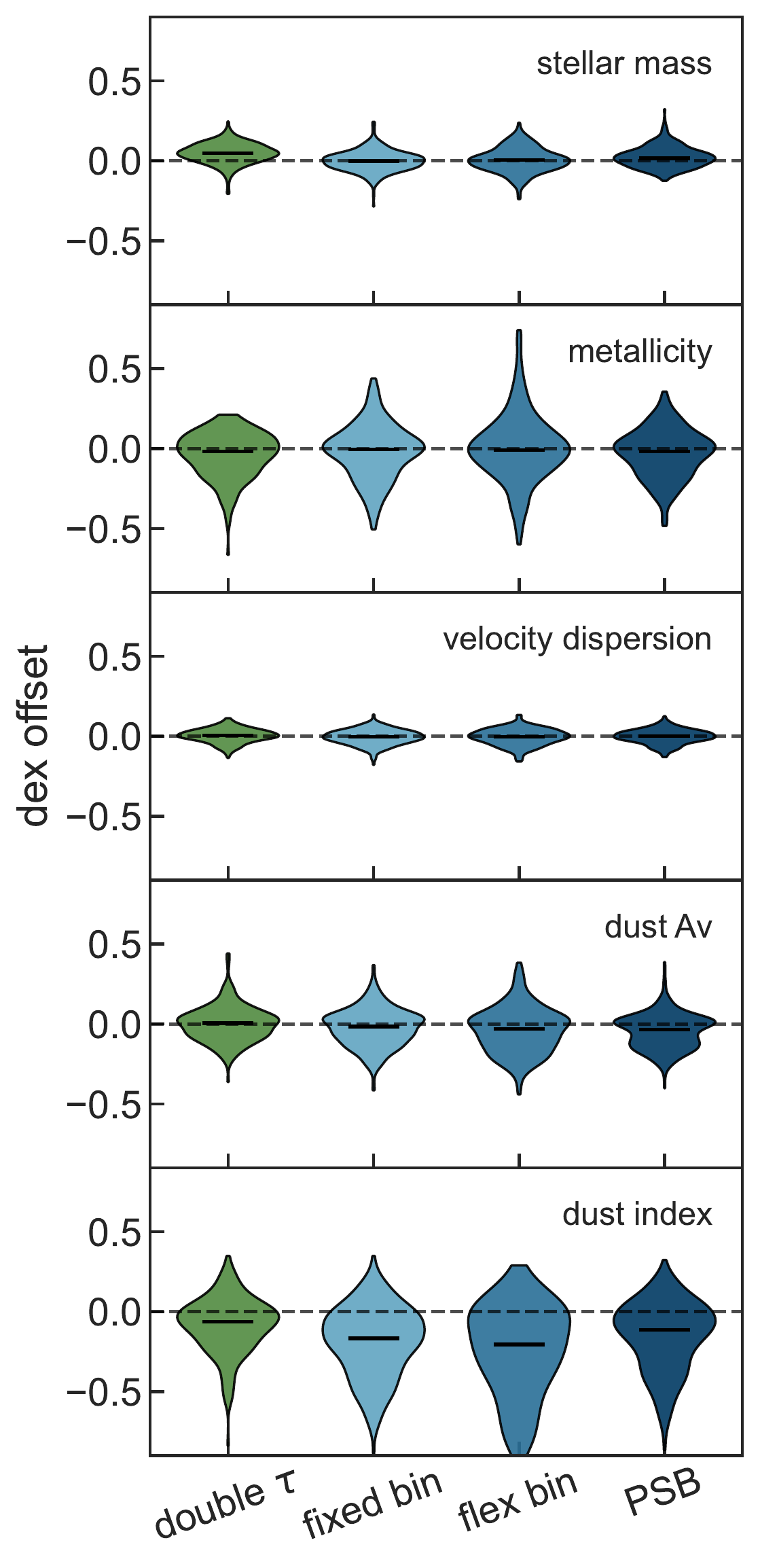}
    \caption{Histograms of the offset between median posterior recovered properties and input properties for 300 mock spectra fit with our four SFH models. In general, these basic properties are recovered with minimal bias and reasonable scatter. Of these quantities, the velocity dispersion is recovered most accurately and the dust index is recovered least accurately. 
    The double-$\tau$ model has the largest stellar mass bias, systematically overestimating $\log{\rm{M}_*/\rm{M}_\odot}$ by 0.05~dex.}
    \label{fig:appendix:hist_recovery}
\end{figure}

In Figure~\ref{fig:appendix:hist_recovery}, we show how well each SFH model is able to recover the basic characteristics of the galaxies: stellar mass, metallicity, dust attenuation, velocity dispersion, dust attenuation, and dust index. All three non-parametric models accurately capture the stellar mass of the galaxy, with offsets of $<0.02$~dex and scatter of $\lesssim0.1$~dex. The double delayed-$\tau$ model slightly overestimates the stellar mass, with a systematic offset of 0.05~dex. 
The metallicities of the galaxies are recovered fairly well by all four models; the scatter is slightly larger than that in the stellar masses at $\sim0.17$~dex, but the median offset is only $0.02$~dex for all three models. 
All four SFH models recover the velocity dispersion of the galaxy both precisely and accurately, with offsets $\le0.02$~dex and scatter $\lesssim0.05$~dex. 

The dust attenuation values A$_v$ are recovered with median offsets $\lesssim0.03$~dex, and scatter of $0.1 - 0.15$~dex.  
The least well-recovered property is the dust index, which has a bias of $0.05-0.21$ and a scatter of $\sim0.25$; all SFH models have a long tail towards underestimated dust indices. This bias is not unexpected: our data are mostly in the rest-frame optical, and do not have much constraining power on the dust index. All mock spectra are generated with a dust index of zero, while our prior range is from -1 to 0.4; therefore, on average our prior pushes us towards lower recovered dust indices than our assumption when generating the mock spectra. Additional data beyond the \squiggle-like spectra and photometry studied here may be required to accurately constrain the dust index.

\subsection{SFRs}
\label{sec:sfrs}

\begin{figure*}[ht]
    \centering
    \includegraphics[width=.75\textwidth]{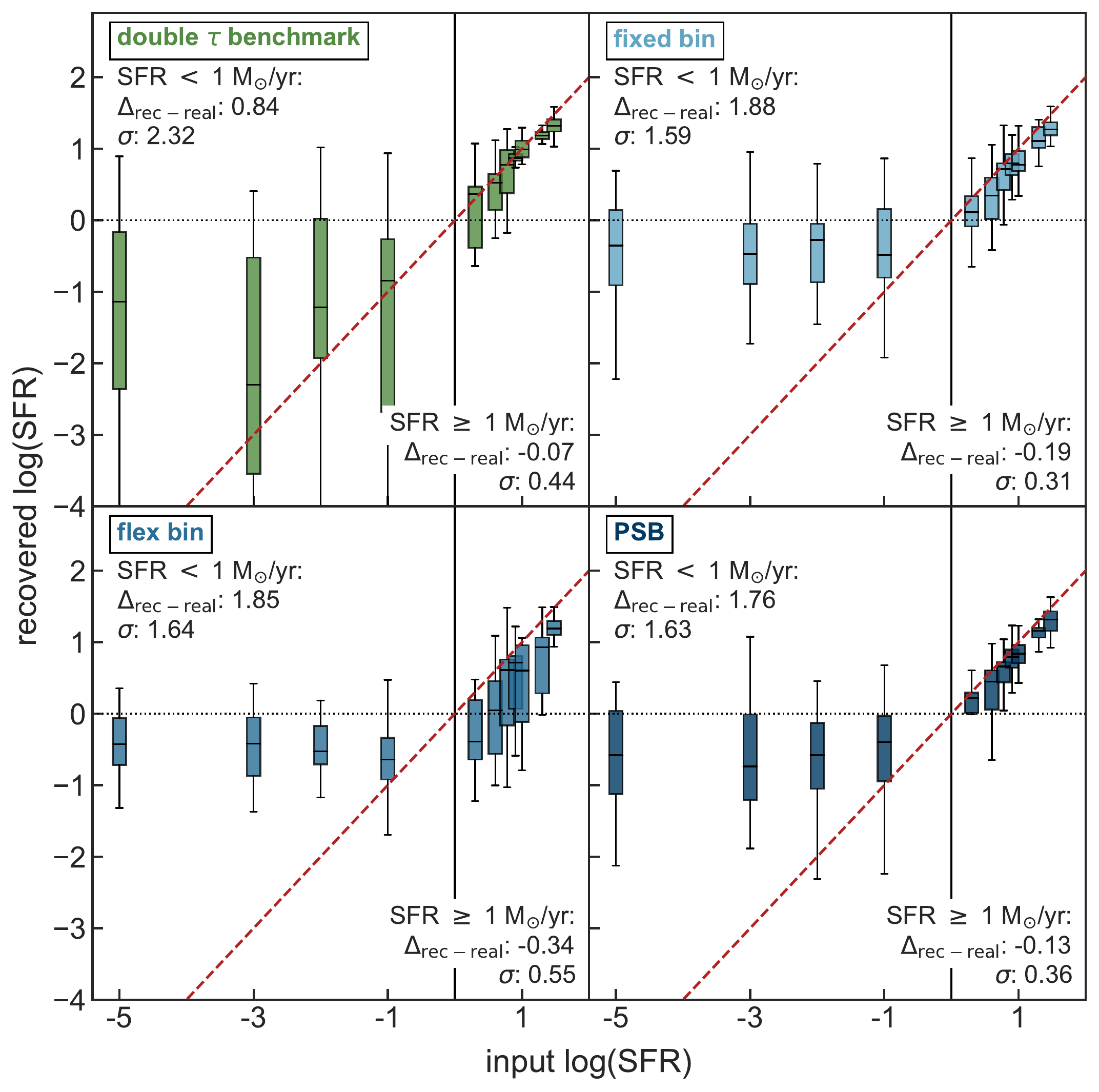}
    \caption{Input and recovered SFRs for 300 mock PSB spectra fit with all four SFH models. Because input SFRs are created on a discrete grid (Section~\ref{sec:mock_generation}), we show a single box-and-whiskers plot for each input SFR value. Each box-and-whiskers represents the median and spread in the recovered SFR values for that input SFR. Text in the upper left and lower right of each panel lists the median offset and scatter between recovered and input SFRs for both low and high input SFRs. While ongoing SFRs are generally recovered well at SFR $>1\rm{M}_\odot$/yr, for all four models SFRs below $1\rm{M}_\odot$/yr are recovered with large error bars and a significant offset towards larger SFR values. For the SDSS-quality data used in this study we recommend treating all SFRs recovered with non-parametric models to be $<1\rm{M}_\odot$/yr as upper limits at $1M_\odot$/yr.}
    \label{fig:sfr_recovery}
\end{figure*}

Next, we test how well our fits are able to recover the ongoing SFR of mock galaxies. Figure~\ref{fig:sfr_recovery} shows the recovered and input log(SFR) for all 300 mock galaxy fits; each panel shows a different SFH model. Because mock galaxies were created on a discrete grid of SFR values (Section~\ref{sec:mock_generation}), for each input SFR we show a box-and-whiskers plot of the median posterior SFRs of all galaxies with that input SFR. 

For all SFH models, the behavior of recovered SFRs differs substantially above and below $\sim1\rm{M}_\odot$/yr. Above $1\rm{M}_\odot$/yr, SFRs are recovered fairly well, with $0.1-0.2$~dex of scatter. All four SFH have recovered SFRs that are biased slightly low, by 0.05~dex for the double delayed-$\tau$ model, 0.19~dex for the fixed bin model, and 0.13~dex for the PSB model. The flexible bin model shows the most bias, systematically underestimating SFRs by 0.34~dex. As explored further in Section~\ref{sec:burst_props}, the relatively large bias in ongoing SFRs for the flexible bin model is likely due to the fact that this SFH parameterization results in all galaxies quenching exactly 100~Myr before observation: the only way the model can produce realistic spectra for galaxies which quenched $>100$~Myr before observation is to underestimate the ongoing SFR.

There is a dramatic shift in how well all SFH models are able to recover low ongoing SFRs. Below $\sim1M_\odot\rm{yr}^{-1}$, the recovered SFR values saturate and the distribution of recovered SFR is flat for all input SFRs. The individual error bars on these overestimated SFRs are relatively large for all models, ranging from $\sim$0.4~dex for the flexible model to $\sim2.4$~dex for the double delayed-$\tau$ model.

Next, we explore {\it why} these SFR floors exist for both the parametric and non-parametric models. To disentangle the effects of the model and prior choices from the effects of the mock data quality, we investigate the sSFR distribution that results from random draws from the priors alone, before the model sees any data. None of the four SFH models directly set a prior on sSFR: however, the priors on the timescale $\tau$ and \logsfrratio\ {\it imply} priors on sSFR. We report the sSFR prior distribution instead of the SFR prior distribution because sSFR is influenced only by the SFH model and priors as listed in Table~\ref{table:sfh_priors}; SFR is also affected by the broad, flat prior on total stellar mass.   
For each model, we take 500,000 calls from each SFH prior assuming a redshift of $z=0.7$. We then calculate the sSFR (in units of SFR$/M_{\rm{formed}}$, which allows us to avoid a time-intensive FSPS call) for each prior draw. 

\begin{figure}
    \centering
    \includegraphics[width=.48\textwidth]{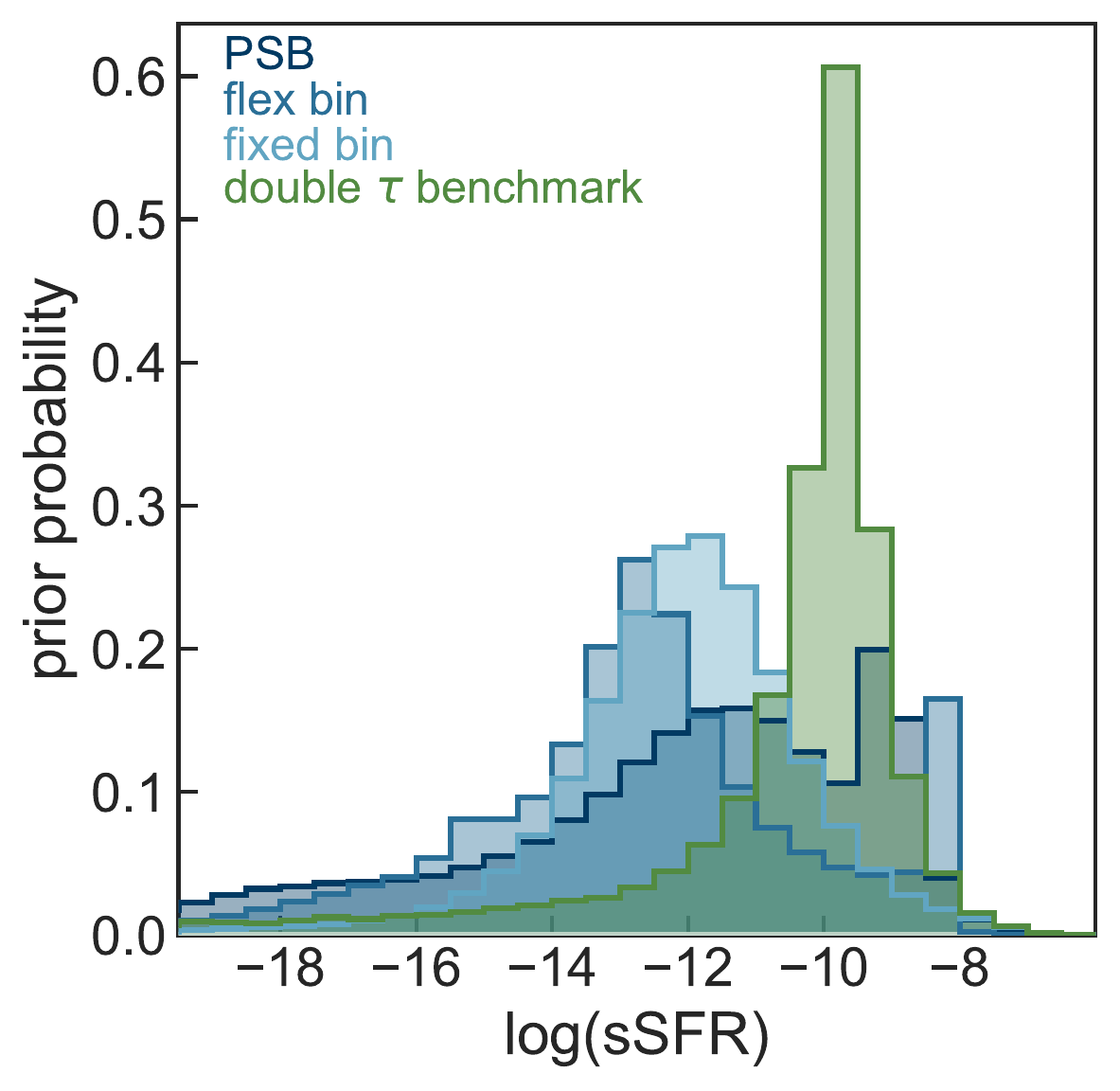}
    \caption{Histograms of the sSFR (SFR/$M_{\rm{formed}}$) resulting from 500,000 draws of the prior distribution for all four SFH models. The three non-parametric models have broad log(sSFR) prior probability distributions centered roughly between $10^{-15}\rm{yr}^{-1}$ and $10^{-10}\rm{yr}^{-1}$. The prior probability distribution for the double delayed-$\tau$ model is much more strongly peaked at $\sim10^{-10}\rm{yr}^{-1}$, but has a long tail towards unphysically low sSFRs of $10^{-300}\rm{yr}^{-1}$.}
    \label{fig:priors}
\end{figure}

We plot a normalized histogram of the sSFR from all 500,000 prior calls in Figure~\ref{fig:priors}. We see that, in the absence of data, all three non-parametric models are able to produce a wide range of sSFRs, including very low sSFRs. The fixed-bin model has the narrowest prior probability distribution, with 16th-84th percentiles from $10^{-13.5}\rm{yr}^{-1}$ to $10^{-10.5}\rm{yr}^{-1}$. The PSB model has the broadest prior probability distribution, with 16th-84th percentiles from $10^{-15.8}\rm{yr}^{-1}$ to $10^{-9.2}\rm{yr}^{-1}$. This indicates that, of the three non-parametric models we test in this paper, the PSB model priors are the least informative of the output sSFR. The fact that all three non-parametric models have significant fractions of their prior probability distribution below $\sim10^{-11}\rm{yr}^{-1}$ suggests that the decrease in accuracy for our SFR mock recovery tests below $\sim1M_\odot$/yr is not due to the non-parametric model and prior choices. Instead, the decreased accuracy at low ongoing SFRs is likely a result of the relatively low S/N of our mock spectra. At such low ongoing SFRs, the differences that slightly different SFRs produce in the spectrum are not visible over the noise. With uninformative data, the prior--- which peaks at $\rm{sSFR}\sim10^{-12}\rm{yr}^{-1}$ for all three non-parametric models--- will dominate the posterior. This also explains the large individual error bars that all three non-parametric models return at low SFRs: the non-parametric models are capable of producing low ongoing SFRs, but the data are simply not constraining for SFR$\lesssim1M_\odot$/yr. For this reason, \citet{suess22} refers to $1M_\odot$/yr as the reliability limit of the SFRs recovered using the PSB SFH model for SDSS-quality spectra: SFRs values below this value should be treated as upper limits at $1M_\odot$/yr. 

{Reaching lower SFR limits with this modeling would require higher S/N spectroscopy or additional wavelength coverage. Spectra covering the H$\alpha$ line in particular would have constraining power on the ongoing SFR: reaching a 5$\sigma$ limit of $0.5\, M_\odot \, \rm{ yr}^{-1}$ at z=0.6 would require reaching depths of $\sim5^{-18}$~erg/s/cm$^2$ across the H$\alpha$ line \citep[][assuming negligible dust attenuation]{kennicutt98}.
This SFR floor of 1 $M_\odot \, \rm{ yr}^{-1}$ is already sufficient to place galaxies an order of magnitude below the star-forming main sequence, which predicts $\sim20\, M_\odot \, \rm{ yr}^{-1}$ of star formation at these masses and redshifts. However, understanding any variations in residual star formation among the PSB sample would benefit from a lower SFR floor: \citet{fumagalli14} finds a typical upper limit for sSFRs in quiescent galaxies of $10^{-11} \, \rm{yr}^{-1}$, our current limiting SFR. For the purposes of our current work, achieving an SFR floor of 1$\, M_\odot \, \rm{ yr}^{-1}$ is sufficient; however, we recommend that future studies carefully consider the effect of model and prior choices on derived SFRs and determine whether their modeling methodology and data quality are sufficient to achieve the desired science goals.}

The double delayed-$\tau$ sSFR prior probability distribution has a median value of $10^{-10.1}\rm{yr}^{-1}$, two orders of magnitude higher than the medians of the non-parametric sSFR prior distributions. The distribution is also much narrower, imposing a much stronger prior on the sSFR. The double delayed-$\tau$ model also has an long low-probability tail that reaches all the way to sSFRs of $10^{-300}\rm{yr}^{-1}$ (compared to minimums of $\sim10^{-30}\rm{yr}^{-1}$ for the non-parametric models). The SFR ``floor" of $\sim10^{0}\rm{M}_\odot\rm{yr}^{-1}$ for our mock recovery tests in Figure~\ref{fig:sfr_recovery}  is definitively below the median value of the prior probability distribution shown in Figure~\ref{fig:priors}. This suggests that our SDSS-quality mock data is sufficiently high quality for the model to determine the best-fit model lies in the tail of the log(sSFR) prior distribution-- just not exactly {\it where} in the tail, given the extremely large error bars in the recovered SFRs and the wide range of prior sSFR probabilities.

\begin{figure}
    \centering
    \includegraphics[width=.48\textwidth]{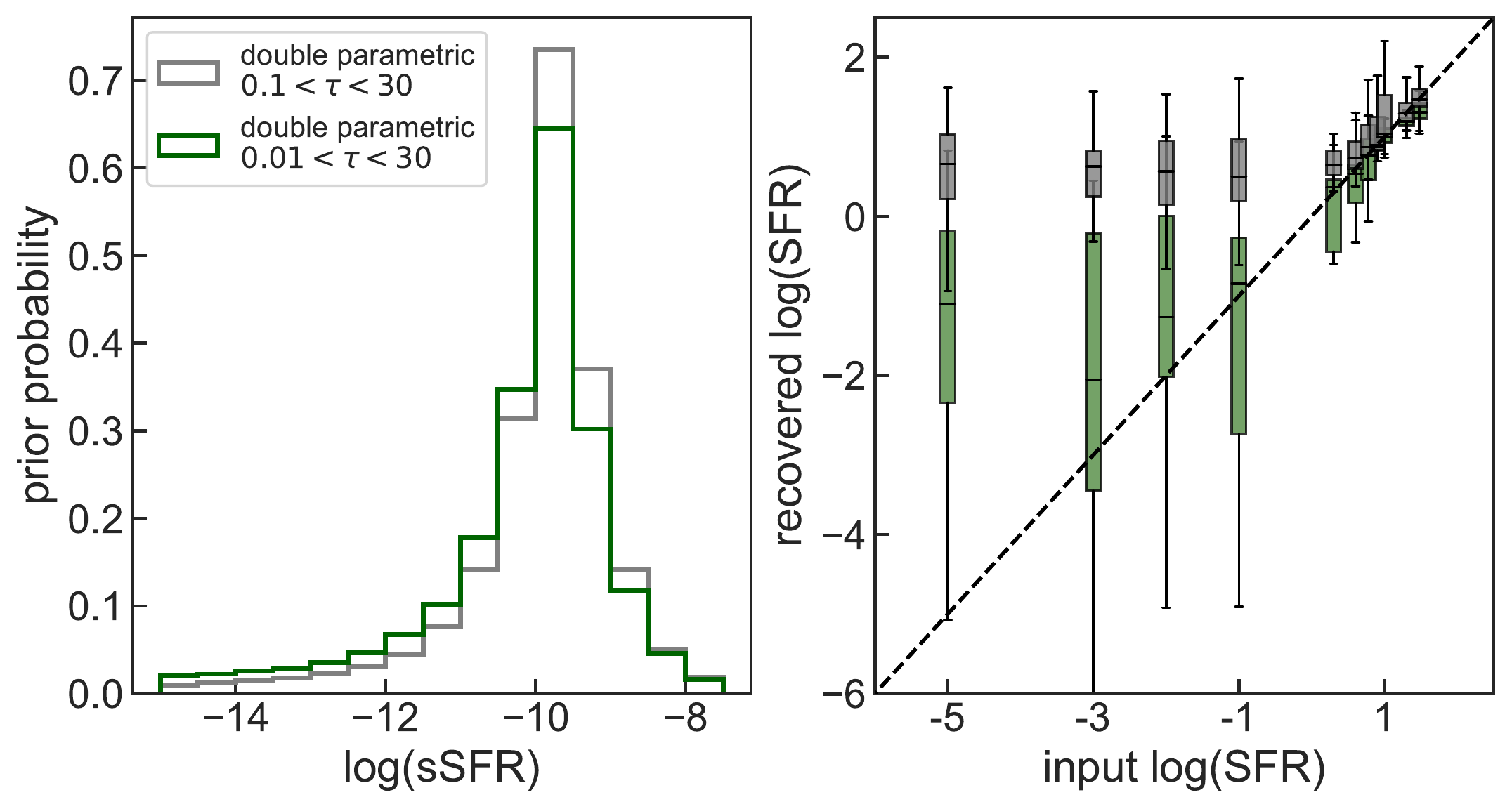}
    \caption{sSFR prior probability distributions (left) and SFR recovery (right) for two different $\tau$ prior ranges for the double delayed-$\tau$ model. Shrinking the prior on $\tau$ by an order of magnitude causes only a small difference in the shape of the prior probability distribution, but changes the inferred SFR for galaxies with ${\rm{SFR}}\lesssim1\rm{M}_\odot$/yr by two orders of magnitude.}
    \label{fig:prior_deltau}
\end{figure}

In Figure~\ref{fig:prior_deltau}, we demonstrate the large impact the $\tau$ prior can have on the SFRs returned by the double delayed-$\tau$ model given that the true sSFRs of our mock galaxies lie in the tail of the prior probability distribution. The left panel shows the sSFR prior probability distribution both for the priors on $\tau_{\rm{young}}$ and $\tau_{\rm{old}}$ used in this paper, which range from $0.01 < \tau < 30$ (green) and a smaller prior ranging from $0.1 < \tau < 30$ (grey). Using a minimum value of $\tau\ge0.1$ is the default in \texttt{Prospector}, and commonly used even in other SED-fitting codes \citep[e.g.,][]{carnall19}. The medians of the two prior distributions differ by only 0.2~dex and are similarly peaked. However, the 16th percentiles of the two prior probability distributions differ by more than four orders of magnitude: $10^{-10.8}\rm{yr}^{-1}$ for the $0.1<\tau<30$ prior, versus $10^{-15.2}\rm{yr}^{-1}$ for the $0.01<\tau<30$ prior. This difference is primarily caused by how much weight the two distributions place very far out in the wings: the minimum sSFR produced by our 500,000 draws of the the $0.1<\tau<30$ prior is $10^{-35}\ \rm{yr}^{-1}$, while the $0.01<\tau<30$ prior produces sSFRs of $10^{-300}\ \rm{yr}^{-1}$. It should be noted that both of these minimum sSFRs are unphysically low. These slight differences in the wings of the sSFR prior probability distribution result mean that a double parametric model using a $0.01<\tau<30$ prior is twice as likely to return $\rm{sSFR}<10^{-11}\rm{yr}^{-1}$ than the $0.1<\tau<30$ prior, and three times more likely to return $\rm{sSFR}<10^{-12}\rm{yr}^{-1}$.

The right panel of Figure~\ref{fig:prior_deltau} shows how well each double delayed-$\tau$ model is able to recover the SFRs of our mock PSBs. Despite a relatively subtle change in the prior probability distributions that occurs mostly in the deeply-unphysical realm of $\rm{sSFR}<10^{-50}\rm{yr}^{-1}$, the recovered SFR values differ wildly. Both models saturate at SFR$\lesssim1M_\odot$/yr, but the $0.1<\tau<30$ prior produces typical SFRs more than two orders of magnitude larger than the $0.01<\tau<30$ prior. \citet{carnall19} shows that changing the shape of the prior distribution on $\tau$ can have similarly large effects on the fits. Figure~\ref{fig:prior_deltau} emphasizes that using a model where the correct solution is in the wings of the prior distribution is not ideal: seemingly small changes in the prior can have very large impacts on the recovered galaxy properties.

\subsection{Mass- and light-weighted ages}
\label{sec:ages}

\begin{figure}
    \centering
    \includegraphics[width=.48\textwidth]{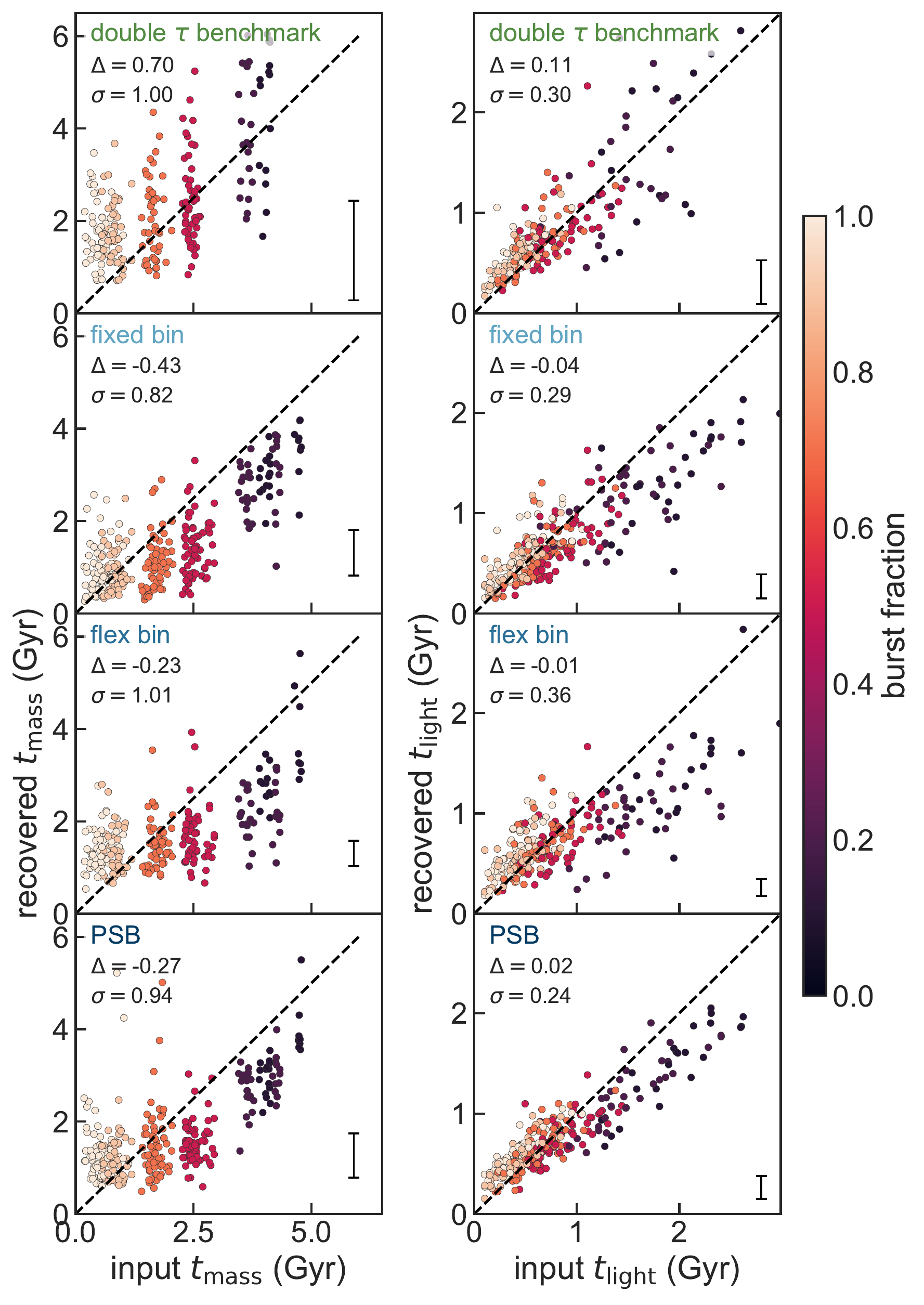}
    \caption{Recovered mass-weighted ages (left) and $r$-band light-weighted ages (right) for each of four SFH parameterizations. Points are colored by the burst mass fraction of the mock galaxy, and a characteristic errorbar is shown in the lower right of each panel. Light-weighted ages are recovered with higher accuracy than mass-weighted ages for all four models. The $\Delta$ and $\sigma$ values reported in each panel show the median offset and scatter (in Gyr) between the input and recovered values. The four models perform similarly well, with the PSB model showing slightly lower scatter in the recovered ages and the double delayed-$\tau$ model showing the highest bias in mass-weighted ages.}
    \label{fig:ages}
\end{figure}

Next, we explore the mass- and light-weighted ages recovered by all four SFH models. We calculate mass-weighted ages directly from the output SFHs. We use \texttt{Prospector} to calculate light-weighted ages for each likelihood draw by setting the FSPS `compute\_light\_ages' flag to True and re-calculating the spectrum without the polynomial calibration factor. We report light-weighted ages averaged between 5580 and 6820$\AA$ (e.g., $r$-band). 

Figure~\ref{fig:ages} shows how well the mass-weighted (left column) and light-weighted (right column) ages are recovered by each SFH model. Each row shows a different SFH model, as indicated by the text in the top left corner of each panel. Data points are shaded by the burst mass fraction of the mock galaxy; galaxies with lower burst fractions have older mass- and light-weighted ages. Light-weighted ages are recovered better than mass-weighted ages by all SFH models. 

The double delayed-$\tau$ model recovers light-weighted ages relatively accurately, with scatter of 300~Myr and a median offset of 100~Myr which is primarily due to overestimates of the light-weighted ages of the youngest galaxies. Mass-weighted ages of the youngest galaxies tend to be very overestimated by this model.

The three non-parametric models have sufficient flexibility to produce a range of mass- and light-weighted ages. The PSB model has slightly less scatter in the recovered light-weighted ages, but only slightly--- the three models perform nearly equally well for recovering mass- and light-weighted ages. All three models recover the light-weighted ages of the youngest galaxies with relatively little bias. However, the mass-weighted ages of young galaxies tend to be overestimated. This indicates that the non-parametric models are forming too much stellar mass at early times. This mass would not contribute significantly to the spectra of these galaxies due to the outshining effect. Therefore, this offset in the mass-weighted ages is likely driven by our prior, which assumes that massive galaxies form relatively large amounts of mass at early times. Both mass- and light-weighted ages of older galaxies tend to be slightly underestimated.

\subsection{Burst and quenching properties}
\label{sec:burst_props}

\begin{figure*}[h!]
    \centering
    \includegraphics[width=.75\textwidth]{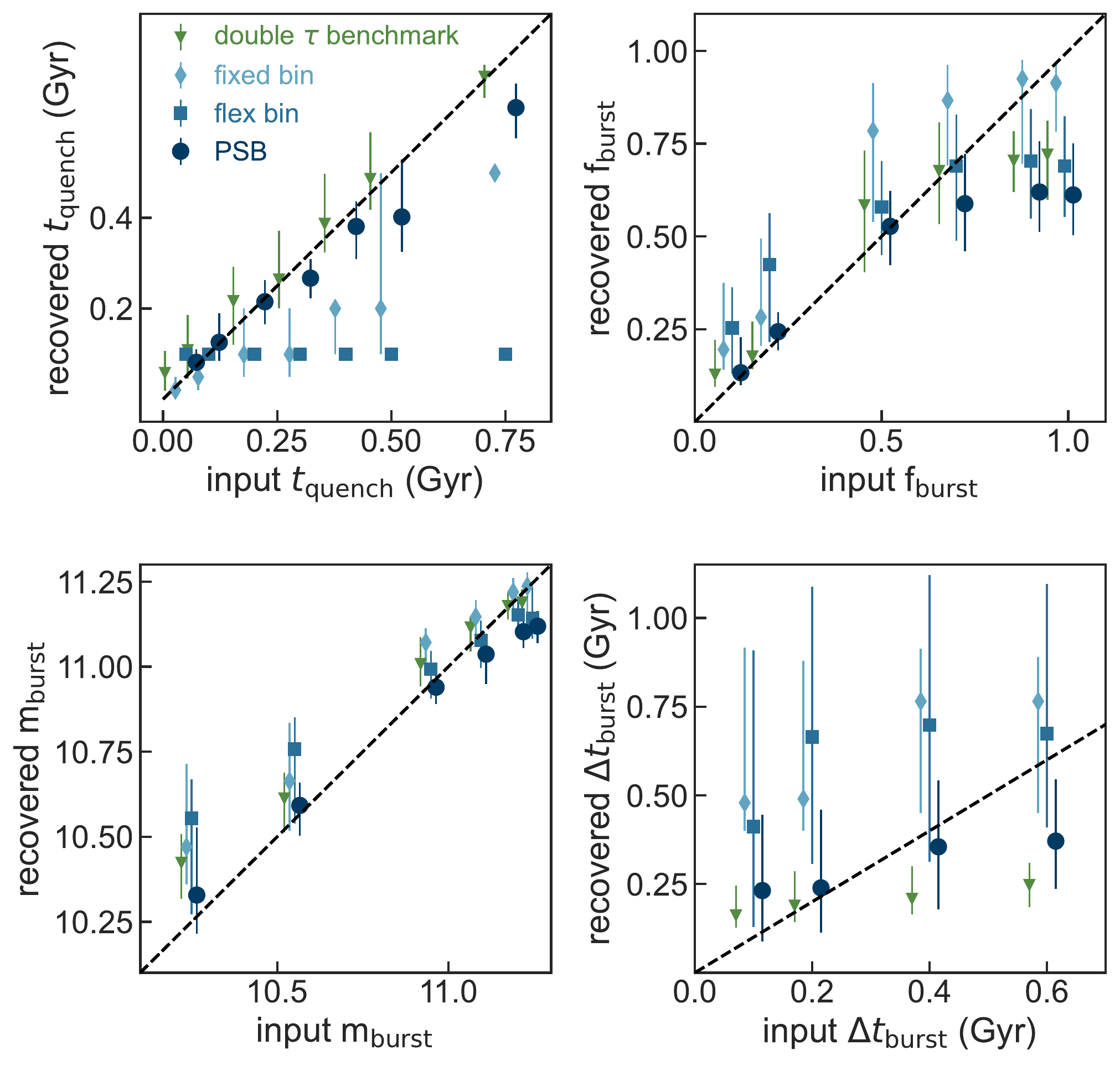}
    \caption{Input and recovered burst properties from fitting 300 mock PSBs with the three non-parametric SFH models. Time since quenching (\tq) is shown in the upper left; burst mass fraction (\fburst) is shown in the upper right; the total mass formed in the burst ($m_{\rm{burst}}$) is shown in the lower left, and the burst duration ($\Delta t_{\rm{burst}}$) is shown in the lower right. The PSB model is the only of the three non-parametric models that is able to accurately recover the key parameter \tq: the flexible-bin model always returns a value of 100~Myr (the width of the final bin), and the median posterior \tq values for the fixed-bin model are both underestimated and show discretization effects related to the choice of bin edges. All three models tend to underestimate high \fburst and $m_{\rm{burst}}$ values, likely due to our conservative choice of priors; the fixed and flexible-bin models also overestimate low \fburst and $m_{\rm{burst}}$ values. None of the three models is able to accurately recover the burst duration, likely because long bursts with lower peak SFR and short bursts with higher peak SFR have the same mass-weighted ages and produce very similar spectra.}
    \label{fig:burst_props}
\end{figure*}

Finally--- and for our purposes, most importantly--- we test how well our fitting is able to recover the properties of the recent burst: when the burst started and ended, and what fraction of the total mass it formed. We note that because these PSBs just quenched by shutting down a burst of star formation, we refer to the end of the burst and the quenching time interchangeably. 

The first challenge in recovering the burst properties is robustly defining ``the burst" in a given output SFH. Previous studies have taken several approaches to defining these quantities. Because \citet{french18} used a double parametric SFH model for their PSB sample, they simply defined the burst as the younger parametric component. \citet{wild20} reported the burst mass fraction as the mass fraction formed in the last 1.0~Gyr, and the quenching time as the time when the galaxy reached 95\% of its total stellar mass. 
Neither of these two burst mass definitions is ideal for our scenario: we would like to define a burst start and end time for both parametric and non-parametric SFHs, ruling out the method used by \citet{french18}. Defining the quenching time as when the galaxy has formed 95\% of its total stellar mass assumes a fixed 5\% of the mass is formed after quenching; because we would like to directly investigate the amount of ``frosting" in observed PSBs, we want our burst definition to be independent of the burst and frosting mass fraction. 

For this work, we choose to define the burst based on the time derivative of the output SFH: the burst begins when the SFR increases sharply, and ends when the SFR decreases sharply. We interpolate each output SFH onto a 100~Myr timegrid, then take the time derivative of the SFR. For the three non-parametric models, we define the burst start as the time when the derivative increases above a threshold value of $100 M_\odot \rm{yr}^{-2}$ and the burst end when the derivative dips below a threshold value of $-100 M_\odot \rm{yr}^{-2}$; this value is tuned by eye using several example fits for all three non-parametric SFHs. The double delayed-$\tau$ model only reaches these thresholds for the very shortest values of $\tau_{\rm{young}}$, resulting in $<5$\% of the fits ``quenching" using this definition. For the double-parametric fits, we thus use a much lower threshold of $\pm10 M_\odot \rm{yr}^{-2}$ to define the start and end of the burst.

Now that we have defined the start and end of the burst, we can investigate how well each SFH model is able to recover the quenching time, burst duration, and burst mass fraction of our mock PSB sample. 
Figure~\ref{fig:burst_props} shows the recovery of the time since quenching (upper left), burst mass fraction (upper right), total mass formed in the burst (lower left), and burst duration (lower right) for each of our four SFH models. Because our mock galaxies were created on a grid of discrete values for these quantities (Section~\ref{sec:mock_generation}), we show one point for each input value, with the value representing the median of all galaxies and the error bar representing the 1$\sigma$ scatter in the recovered quantities.  

The top left panel of Figure~\ref{fig:burst_props} shows that the double delayed-$\tau$ model recovers quenching time well, with just 90~Myr of scatter. However, there are dramatic differences in how well each of the three non-parametric SFH models are able to recover \tq, our primary burst quantity of interest. The first ``out-of-the-box" model, the fixed bin model, is only able to return specific values for \tq: because the SFR can only change {\it at} the pre-specified bin edges, the model must quench at one of these bin edges. Given our choice of bin edges, this means the SFH can quench at 20, 50, 100, 200, or 500~Myr before observation. We clearly see this discretization in the recovered \tq values: the youngest galaxies are recovered with 20~Myr \tq values, then 50~Myr~\tq values, and on up. However, the jumps in the recovered \tq values do not translate perfectly to the input \tq values: for longer input \tq, the fixed bin model significantly underestimates the time since quenching. This results in the fixed-bin model underestimating \tq by $\sim100$~Myr on average, with a scatter of 120Myr. Additionally, the error bars on recovered \tq values for the fixed-bin model are quite large because they are proportional to the bin spacing. We expect that the bias and scatter in the recovered \tq values would decrease if the number of timebins were significantly increased, because there would be a larger set of allowed \tq values. However, increasing the number of bins beyond the current value of 9 significantly increases the required computational time for the fit: in our testing, single-core fits with $>9$ bins hit the maximum cluster wall clock time (72hr) before converging. 

The second ``out-of-the-box" model, the flexible-bin model, returns a \tq value of exactly 100~Myr no mater the input \tq value. This value is both precise (the 1$\sigma$ error bars are equal to zero) and completely uncorrelated with the actual quenching timescale of the mock galaxy. This occurs because each flexible bin forms an equal stellar mass. 
As discussed in detail in \citet{leja19a}, this imposes a minimum floor on the allowed sSFR in the flexible bins that is too high to be considered ``quenched" by our definition. Therefore the flexible-bin model quenches at exactly the transition between the flexible bins and the final fixed-edge bin, no matter what length we choose for the final bin. Because this SFH model cannot constrain the quenching time, it is unsuitable for use with PSBs.

The third non-parametric model, the PSB model, is able to recover the \tq values of the input galaxies with much higher accuracy than either the fixed or flexible-bin models. \tq is slightly underestimated for galaxies which quenched $>400$~Myr before observation, but overall the \tq values are recovered with an average offset of just 10~Myr and a scatter of 90~Myr. This increased accuracy is because the length of the final fixed-edge bin is a free parameter in the fit. This means that, unlike either of the two ``out-of-the-box" models, the PSB model can produce arbitrary \tq values that are informed by the data, not the way we choose to model the SFH.

Figure~\ref{fig:burst_props} shows that the {\it absolute} amount of mass formed in the burst is relatively well-recovered by the PSB and double delayed-$\tau$ SFH models. There is a small offset between the recovered and input burst mass values, $\sim0.06$~dex, driven primarily by underestimated burst masses at the high-mass end. However, the {\it fraction} of mass formed in the recent burst has nearly double the scatter and offset as the absolute burst mass. In particular, there is an obvious offset between the recovered and input burst mass fractions at the high f$_{\rm{burst}}$ end. These high burst fractions represent very extreme SFHs, where 90-99\% of the galaxy's total mass was formed in the recent burst. Even with these high input burst fractions, the recovered SFHs appear to saturate at $\sim80$\% of the mass formed in the recent burst. This is likely due to our continuity SFH prior, which is centered around the average SFH of a UniverseMachine quiescent galaxy (Section~\ref{sec:sfh_priors}): forming just 1-10\% of the total mass of the galaxy before the last $\sim500$Myr is unlikely given this prior. Because the total burst mass is recovered with higher accuracy than the burst mass fraction, this is likely an outshining problem: for high burst masses, our SFH prior allows the fits to ``hide" a relatively large number of old stars under the large recent burst. Whether such extreme SFHs actually exist in practice for massive galaxies at these intermediate redshift ranges is unclear. 
The fixed- and flexible-bin models tend to overestimate both f$_{\rm{burst}}$ and m$_{\rm{burst}}$ for low burst fractions; like the PSB model, they underpredict f$_{\rm{burst}}$ for the most bursty mock galaxies. This is likely correlated with the underestimated \tq values for the fixed- and flexible-bin models: because these models tend to quench later, they form more stars at late times and have higher recovered burst masses and fractions.

None of the four SFH models is able to accurately constrain the duration of the burst. The double delayed-$\tau$ model can only produce rapid quenching events if the timescale of the recent burst is extremely short, $\tau_{\rm{young}}\lesssim 100\rm{Myr}$. For longer bursts, the model must either match the longer star formation timescale {\it or} the rapid SFR dropoff at quenching. As a result, all of the double delayed-$\tau$ fits that are identified in Figure~\ref{fig:burst_props} as quenching do so very rapidly, and the recovered burst duration is flat at $\sim100\rm{Myr}$ no matter the input value. 
Figure~\ref{fig:burst_props} shows that the scatter between recovered and input burst duration for the three non-parametric models is 250~Myr; the error bars on individual recovered measurements are similarly high. There is little correlation between the input and recovered burst duration. This may indicate that our data quality and modeling is insufficient to distinguish between a short burst with high peak SFR and a longer burst with lower peak SFR.

\subsection{Bayesian model selection}
One of the advantages of sampling with dynamic nested sampling codes such as \texttt{dynesty} is that they directly compute the Bayesian evidence $Z$. This allows us to compute the Bayes factor evidence and quantify whether our mock data is better fit by the PSB SFH model or one of the other four SFH models. \citet{kass95} suggest computing the Bayes factor as $B\equiv 2\ln{(Z_1/Z_2)}$: $B>10$ indicates that the data has a very strong preference for model 1, $6<B<10$ indicates a strong preference for model 1, $2<B<6$ indicates a preference for model 1, and $0<B<2$ indicates a weak preference for model 1 that is ``not worth more than a bare mention." In this formulation, negative values of $B$ indicate a preference for model 2 over model 1. \citet{lawler21} suggest that this method of Bayesian model selection is able to successfully determine the ``more correct" SFH model given sufficiently high S/N.

We compute the Bayes factor evidence for each SFH model compared to our PSB SFH model. In all cases, at least 99\% of the mock spectra very strongly prefer the PSB SFH model. We find that 297 mock spectra very strongly prefer the PSB SFH model over the double delayed-$\tau$ model; the remaining three prefer or strongly prefer the double delayed-$\tau$ model. 297 mock spectra very strongly prefer the PSB SFH model over the fixed-bin SFH model; the remaining three spectra have a strong or moderate preference for the PSB SFH model. 298 spectra have a strong preference for the PSB SFH model over the flexible-bin model; one has a weak preference for the flexible-bin model, and one has a strong preference for the flexible-bin model. These results bolster our findings in Section~\ref{sec:burst_props}: increased flexibility in the recent SFH shape means that the PSB SFH model almost always provides better fits to the mock spectra than any of the other three SFH models we test in this paper.

\subsection{Summary: common failure modes for SFH models}

\begin{figure*}[ht]
    \centering
    \includegraphics[width=.95\textwidth]{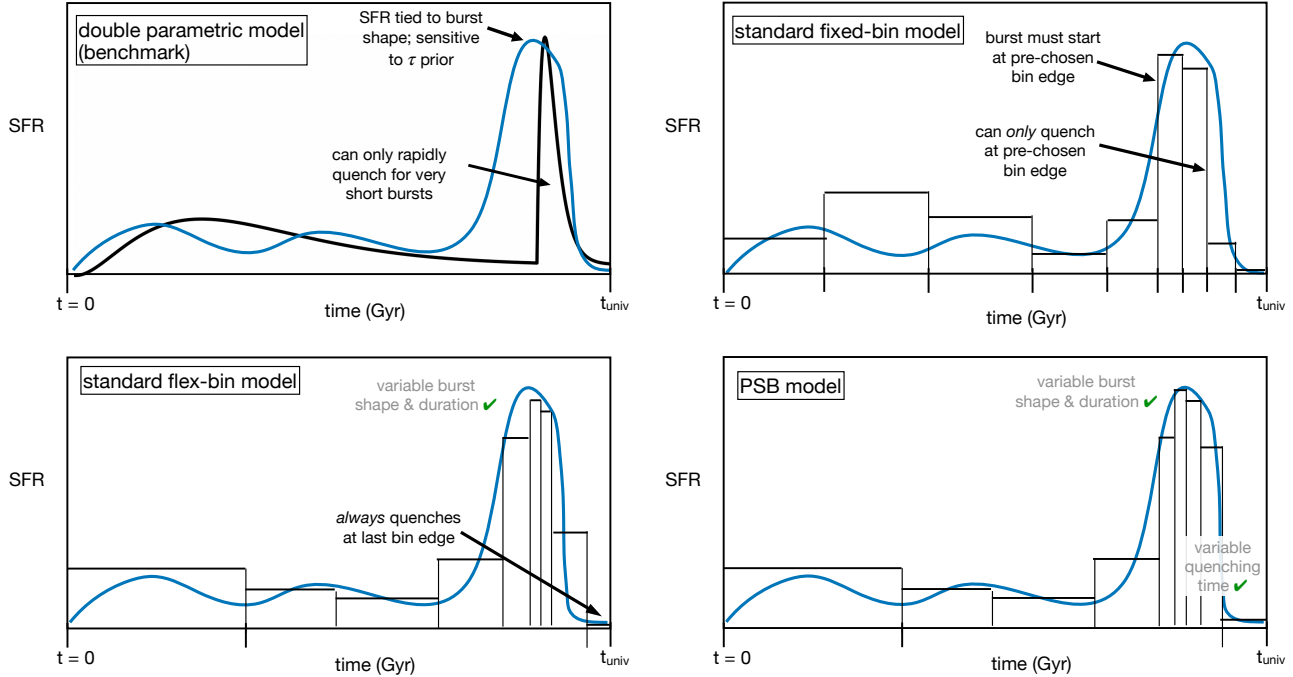}
    \caption{Common failure modes for the four SFH models tested in this paper. The prescribed shape of the double delayed-$\tau$ model ties the burst shape to the ongoing SFR, and only allows for rapid quenching to take place for very short bursts. The standard fixed-bin non-parametric model can only change SFR at the pre-chosen bin edges; this introduces error into the recovered burst start time, burst shape, and quenching time. The standard flexible-bin non-parametric model always quenches at the last bin edge; the flexible bins form too much stellar mass to ever be considered quenched. Only the PSB model allows for variable burst shape, duration, and quenching time.}
    \label{fig:failure_modes}
\end{figure*}

Figure~\ref{fig:failure_modes} shows a cartoon visualization of the most common pitfalls of the four SFH models we test in this paper. 

The strict form of the double delayed-$\tau$ model causes difficulties even in this ideal test case, where the shape of the mock galaxy SFH is very similar to the parametric model. The ongoing SFR is tied to the burst shape: this means that the output SFR is very sensitive to the exact prior used on the star formation timescale $\tau$. As discussed in Section~\ref{sec:sfrs}, allowing low $\tau$ values of 0.01~Gyr$^{-1}$ is necessary to achieve low ongoing SFRs. However, these low $\tau$ values also place an uncomfortably large amount of probability at unphysically low sSFRs of $<10^{-50}\rm{yr}^{-1}$. Furthermore, this model can only quench rapidly for extremely short values of $\tau$. Longer bursts or multiple bursts cannot accurately be modeled with this parametric form.

In the non-parametric fixed-bin model, the SFR can only change {\it at} the bin edges. As a result, the burst can only begin and end at a bin edge. This means that recovered quenching times are always exactly equal to one of the pre-chosen bin edges. Even in the best possible case, this produces an expected error on \tq of half the bin spacing. As discussed in Section~\ref{sec:burst_props}, this effect would be minimized as the number of time bins is increased and there are more discrete quenching times available to the model. However, adding even more bins to the SFH rapidly becomes computationally infeasible.

We find that the non-parametric flexible-bin model {\it always} quenches at the last bin edge. This is due to model construction: each flexible bin forms an equal amount of stellar mass. To understand how this translates to SFRs, we turn to the \squiggle PSB sample described in \citet{suess22}. The \squiggle PSBs have total masses of $\sim10^{11.25}M_\odot$, and the majority of galaxies have burst mass fractions of at least 25\%. With five flexible-edge bins, each bin thus forms $\gtrsim10^{10}M_\odot$. Reaching an SFR of $\sim10M_\odot$/yr would require that the last flexible bin be a full gigayear long-- far less than the \tq values expected for post-starburst galaxies. The exact value of the minimum SFR floor in the flexible bins depends on the specific galaxy, but the general picture holds: numerically, the standard flexible-bin model {\it cannot} quench before the last bin, no matter what value is chosen for the final bin edge. Out of the box, the flexible-bin model is thus unsuited for recovering the properties of PSBs.

Our PSB SFH model was designed to avoid these common failure modes. By including the length of the last bin as a free parameter in the fit, we avoid the issues that both the fixed and flexible bin models have recovering \tq: \tq is not forced to be equal to some pre-chosen value or set of values, but can be directly informed by the data. Parameterizing the pre-quenching SFH using flexible bins also allows for the burst shape to be free. This model has been added to the public distribution of \texttt{Prospector} as the ``continuity\_psb\_sfh" template; the number of fixed bins as well as the number of flexible bins can be modified by the user.

\section{Testing the PSB SFH model on quiescent and star-forming galaxies}
\label{sec:sfqui}

\begin{figure*}[ht]
    \centering
    \includegraphics[width=.95\textwidth]{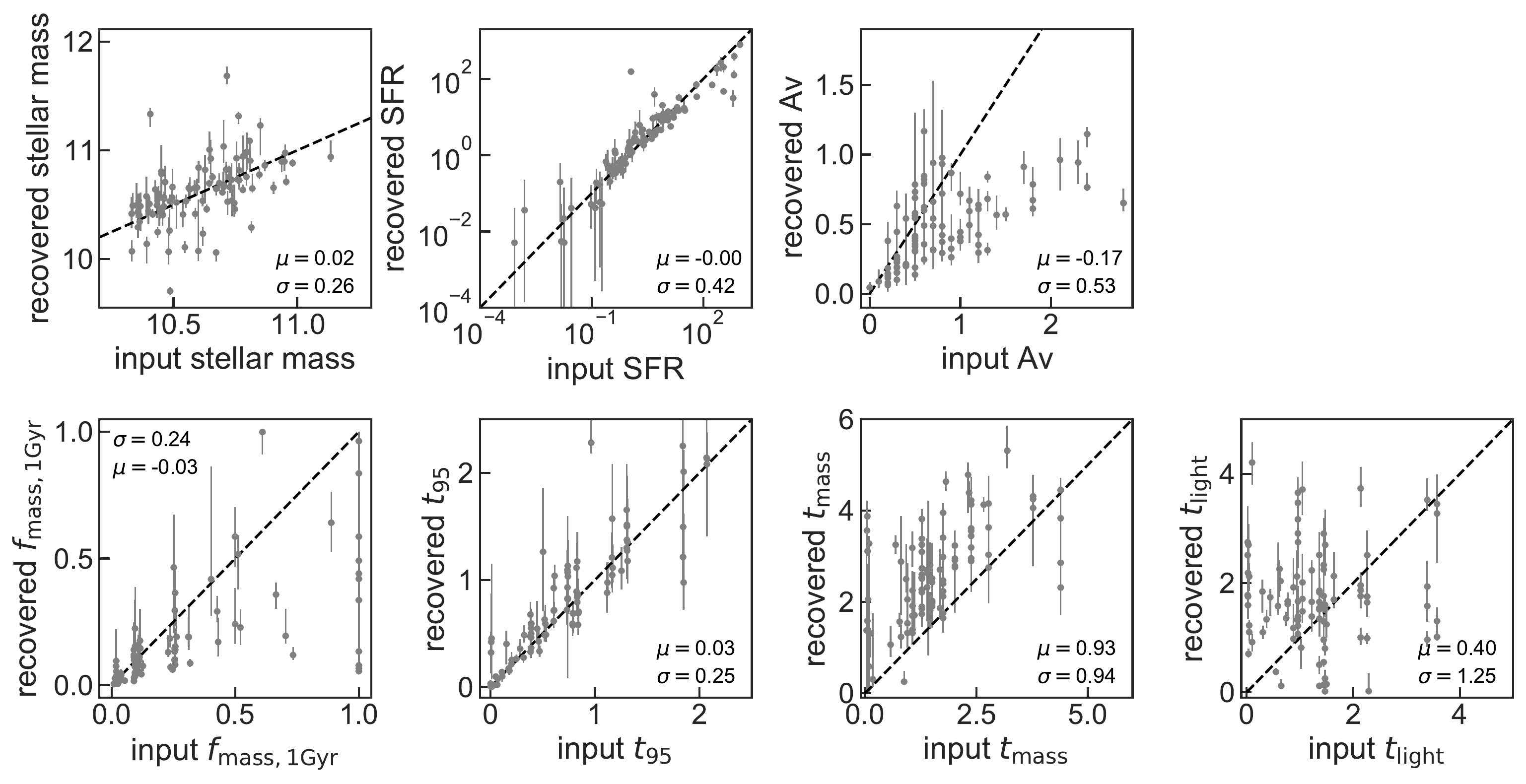}
    \caption{Recovered parameters as a function of input parameters for the PSB SFH model tested on mock star-forming and quiescent galaxies. Panels show the stellar mass, the ongoing SFR, the dust attenuation $A_v$, fraction of the total mass formed in the last 1.0~Gyr ($f_{\rm{mass, 1Gyr}}$), how long ago the galaxy formed 95\% of its current stellar mass ($t_{95}$), the mass-weighted age ($t_{\rm{mass}}$), and the light-weighted age ($t_{\rm{light}}$). Stellar masses, ongoing SFRs, and $t_{95}$ are recovered accurately. The scatter in recovered SFRs increases below $\sim1M_\odot$/yr; as for PSBs (Figure~\ref{fig:sfr_recovery}), this is likely because very low ongoing SFRs are not distinguishable from noise in these SDSS-quality mock spectra. $t_{\rm{mass}}$ tends to be overestimated and $f_{\rm{mass, 1Gyr}}$ tends to be underestimated, indicating that the SFH fits form more mass at early times than these mock galaxies. This is expected, as the mocks are created with a simple delayed-$\tau$ SFH that does not include star formation at early times, while the PSB SFH model prior does assume early star formation.}
    \label{fig:sfqui}
\end{figure*}

In Section~\ref{sec:pick_model}, we identified the best non-parametric model for describing the SFHs of PSBs. Here, we validate that our PSB SFH model has sufficient flexibility to also describe the SFHs of both star-forming and quiescent galaxies. Our goal is to show that the PSB SFH model is suitable for general use where the galaxy type is not necessarily known in advance of fitting.

Figure~\ref{fig:sfqui} shows how well the properties of the mock star-forming and quiescent galaxies described in Section~\ref{sec:mock_sfqui} are recovered by \texttt{Prospector} using the PSB SFH model. Stellar mass is recovered well, with $0.25$~dex of scatter. The ongoing SFR is also recovered well, with no significant offset and a scatter of $\sim0.4$~dex. The scatter in the recovered SFRs increases significantly below $\sim0.1M_\odot$/yr. This limit is a factor of $\sim10$ lower than the SFR reliability limit for the mock PSBs (Figure~\ref{fig:sfr_recovery}), likely because the SFHs of the mock quiescent galaxies are changing less rapidly and are easier for the model to reproduce.
Like for the mock PSBs, the existence of an SFR reliability limit for these recovery tests is likely because very low ongoing SFRs do not appreciably change the SDSS-quality spectrum. Returned SFRs are strongly influenced by the prior, which peaks at sSFR$\sim10^{-12}\rm{yr}^{-1}$. Critically, we note that the PSB SFH model is able to accurately reproduce a wide range of ongoing SFRs: while the model was developed to accurately reproduce recently-quenched galaxies, the model is able to return both star-forming and quiescent solutions. We see that $A_v$ values tend to be underestimated by $\sim0.2$~mag, especially for high input $A_v$ values. This may partially be due to the fact that our fitting includes a free dust index which is not well-constrained, but the mock spectra are all generated with a \citet{calzetti00} dust law. We note again that the tests performed in this paper are not designed to investigate the most appropriate dust law to use in SED fitting; fitting with a variety of different dust laws may increase the scatter in recovered SFRs and $A_v$ values, especially for dusty star-forming galaxies. Use of a non-uniform dust screen model may also improve how well SED fitting is able to recover the dust attenuation law and SFR \citep{lower22}. 

The bottom row of Figure~\ref{fig:sfqui} shows three different probes of the SFH. Because these star-forming and quiescent galaxies did not necessarily experience recent starbursts, we do not show \tq and \fburst as in Section~\ref{sec:pick_model}. Instead, $t_{95}$ shows the lookback time when the galaxy formed 95\% of its stellar mass, $t_{50}$ shows the lookback time when the galaxy formed 50\% of its stellar mass, and $f_{\rm{mass,1Gyr}}$ shows the fraction of the total mass formed in the last 1.0~Gyr. $t_{95}$ is recovered accurately, with 0.25~dex of scatter. This indicates that the recent star formation activity in these galaxies is recovered well. However, we see that $t_{50}$ tends to be overestimated and $f_{\rm{mass,1Gyr}}$ tends to be underestimated; this indicates that the recovered SFHs form more mass at early times than the mock galaxies. This is not surprising: the mock galaxies were created using delayed-$\tau$ SFHs, which have no star formation before the current episode. The SFH prior, in contrast, forms a significant amount of mass at early times. Because of the outshining problem, this prior is the primary determinant of the early-time SFH.

\section{Discussion: how to choose the right SFH model}
The majority of this paper focused on mock recovery tests specifically designed to test how well different SFH models are able to recover the properties of mock PSBs. PSBs are interesting in their own right: understanding the SFHs of these recently-quenched galaxies can provide estimates of their burst mass fractions, average ages, and time since quenching. These quantities can be used to compare to theoretical quenching processes and to help understand how these galaxies evolve after shutting down their star formation. Insights from these mock tests can relatively easily be applied for SED modelers seeking to understand the SFHs of recently-quenched galaxies. However, these tests are also more broadly applicable: PSBs represent an extreme use case to fully test the accuracy of SFH modeling. Their large recent bursts, sharp quenching events, variable ongoing SFRs, and possible multiple episodes of star formation push SFH models to the limit. Here, we consider how the lessons learned from fitting these extreme galaxies with different SFH models can be applied more generally for a wide range of SED fitting use cases. 
\subsection{Choosing a model to recover basic quantities}
``Basic quantities" derived from SED modeling are relevant for a wide variety of use cases, and include stellar mass, metallicity, dust content, SFR and sSFR, and average age. 
A key insight of the tests we perform in Section~\ref{sec:pick_model} is that all three non-parametric SFH models have sufficient flexibility to accurately recover all of these basic quantities. This suggests that any ``out-of-the-box" non-parametric model is sufficient for general SED fitting. 
In Figure~\ref{fig:appendix:hist_recovery}, we show that all three non-parametric models perform equally well at recovering the stellar mass, metallicity, velocity dispersion, $A_v$, and dust index of mock PSBs. The SFR recovery shows slight differences between the models--- e.g., in Figure~\ref{fig:sfr_recovery}, we show that the standard flexible-bin model has an additional 0.1~dex systematic offset in recovered SFRs compared to the standard fixed-bin model and the PSB model--- but in broad strokes, all three models perform quite similarly. Figure~\ref{fig:ages} shows that the three non-parametric models have slight differences in the scatter between input and recovered mass- and light-weighted ages, but again the models perform nearly interchangeably. This finding indicates that, when seeking to recover only these basic quantities, essentially {\it any} of the non-parametric SFH models tested here is sufficient. This does of course come with caveats: no SFH model is able to uncover information that is beyond the limits of the data, and when the data are uninformative the prior distribution has a significant impact on the posteriors. In these cases, caution should be used when comparing the results of SED fitting performed using different prior assumptions. These results also apply to the distributions of recovered quantities for a sample of several hundred galaxies: the SFRs and ages of a single galaxy may differ when when a different SFH model is used to perform the fitting. But overall, our results indicate that for general SED fitting, any non-parametric model has sufficient flexibility to accurately recover the basic properties of a sample of galaxies. 

We find that the double delayed-$\tau$ model is also able to accurately reproduce the basic properties of our mock PSBs. However, we stress that our mock PSBs are the ``best case" scenario for testing this model: the input SFHs have a very similar functional form to the double delayed-$\tau$ model, and parametric models can only accurately reproduce results if the correct answer is contained within the model space. In this paper, we use relatively simple mock SFHs composed of an older delayed-$\tau$ component as well as a recent tophat burst. The double delayed-$\tau$ model is able to recover this functional form only if the timescale $\tau$ is very short. Figure~\ref{fig:prior_deltau} demonstrates that changing the $\tau$ prior to exclude timescales $0.01 \le \tau \le 0.1$ can bias the recovered SFRs by two orders of magnitude. \citet{lower20} shows that stellar masses can also be biased if the true SFH does not perfectly align with the functional form of the parametric model. These model mismatches can be difficult to identify: if a given model is unable to access the ``true" region of parameter space, incorrect values can be returned with drastically underestimated error bars (as seen in Figure~\ref{fig:prior_deltau}). This highlights one of the dangerous pitfalls of parametric SFH models: basic quantities can be highly biased {\it without the user being able to tell from the estimated uncertainties}. Due to this issue, caution should be used when interpreting the results of SED fitting using parametric models. When using parametric models, scientific conclusions should always come with a discussion of which parts of parameter space are excluded by the SFH functional form.

\subsection{Choosing a model to recover higher-order quantities}
While all three non-parametric models are able to recover basic SED fitting quantities, they show significant differences in performance when attempting to recover higher-order SFH quantities such as \tq and \fburst. As shown in Figure~\ref{fig:burst_props}, neither ``out-of-the-box" non-parametric model is able to constrain \tq. The flexible-bin model always quenches at the final bin edge, and the fixed-bin model systematically underestimates \tq. Error bars on \tq for the fixed-bin model reflect the chosen bin spacing, not the ability of the data to constrain \tq; error bars on \tq for the flexible-bin model are equal to zero, because the model is unable to vary this quantity. The three models also have differences in the recovered \fburst value, again driven by differences in model flexibility as opposed to differences in the data. Accurately constraining \tq and \fburst required building a new SFH model specifically designed and tested to recover these higher-order SFH quantities. 

These results indicate that, if the user is attempting to recover specific higher-order SFH quantities--- e.g., \tq, \fburst, how long it took for a galaxy to quench, the length of a recent starburst, the fraction of the total mass formed within a specific time interval, the timescale on which SFR variability occurs--- it is {\it essential} to carefully consider the priors and the SFH model flexibility. These higher-order quantities can be significantly affected by relatively small choices made during SFH model construction. Mock recovery tests are critical to disentangle the effects of the model and priors from the scientific results.

With this point in mind, we note that the PSB SFH model designed in this paper is simply a slightly more flexible version of existing ``out-of-the-box" non-parametric models. We added a single additional free parameter, the width of the final timebin, and allow for the user to set the total number of fixed and flexible bins. This means that the PSB SFH model has increased flexibility in the most recent part of the SFH, exactly the portion of the SFH which is most constrained by the data. This additional flexibility is broadly applicable to a large variety of use cases where the recent SFH varies on rapid time scales. Beyond modeling quenching events, \citet{chaves20} suggest that the effect of SFH on galaxy color is almost entirely driven by the fraction of the mass formed in the past 1~Gyr: accurately recovering this recent SFH is critical to accurately recover the colors and physical properties of all galaxies. While additional tests of the model should be performed when using it to recover higher-order quantities that are not detailed in this paper, for most use cases we suggest that there are few downsides to using this SFH model over a different non-parametric SFH model--- and even a potential upside, of obtaining more information about the most recent SFH. Our PSB SFH model is available in \texttt{Prospector} as ``continuity\_psb\_sfh" in the template library.

\subsection{Caveats \& future work}
While this work represents a first step towards understanding how to best use non-parametric SFH models to understand the quenching process, many open questions remain. Our mock observations for this work consist of SDSS-quality spectra intended to be directly analogous to the \squiggle survey of intermediate-redshift PSBs. While these mock data are similar in quality to what may be expected from the upcoming spectroscopic surveys such as DESI, PFS and MOONRISE, the details of how well burst and quenching properties can be recovered may differ for purely photometric data, such as that expected from the upcoming {\it James Webb Space Telescope}. Furthermore, the mock observations we use are geared specifically towards understanding the properties of PSBs, and our mock SFHs are a relatively simple model of a tophat burst on top of an older delayed-$\tau$ component. Future mock recovery tests using the SFHs of simulated galaxies \citep[e.g.,][]{smith15,guidi16,iyer20} may provide additional insights into the best SFH models to recover galaxy properties.  

We also note that both the mock galaxies and our SED fitting models were generated with FSPS, and use the same underlying stellar isochrones and spectral libraries. These tests are thus insensitive to any possible differences between these models and true galaxies caused by binary stars \citep[e.g.][]{eldridge17} or TP-AGB stars \citep[which may be especially important in PSBs, e.g.,][]{kriek10}.

\section{Conclusions}

In this paper, we explore how well different SFH parameterizations are able to recover the properties of mock PSBs. We test one parametric SFH model as well as three ``non-parametric" SFH models. We create mock PSBs with known stellar populations and SFHs based on the properties of observed intermediate-redshift PSBs from the \squiggle survey \citep{suess22}. We then fit these mock observations with the \texttt{Prospector} SED fitting code \citep{johnson20} to test how well each SFH model is able to recover the known properties of each mock galaxy. 

We find that the double delayed-$\tau$ model is able to accurately reproduce the stellar masses and SFRs of our mock PSBs as long as very short $\tau$ values are allowed by the model. This model is also able to accurately reproduce quenching times. However, galaxies are only identified as ``rapidly quenched" if the timescale $\tau$ is $\lesssim100$~Myr. Because of this degeneracy between the burst duration and the speed at which star formation shuts off, this parametric model is unable to recover the duration of the recent burst. The prior extending to low $\tau$ values required to accurately reproduce SFRs and quenching events also places a large amount of probability at unphysically low ongoing sSFRs of $10^{-300}$ to $10^{-50}\rm{yr}^{-1}$. This tension does not exist for more flexible non-parametric models.
The recovery tests in this paper represent a nearly-ideal case where the true SFH and the parametric model have very similar forms. The double delayed-$\tau$ model may not have sufficient flexibility to recover more complex input SFHs.

All three of the non-parametric SFH models we test describe the SFH as a piecewise function where the SFR varies between different time bins. In the fixed-bin model, the edges of the bins are set by the user and do not change. In the flexible-bin model, the edge of the first and last bin remain fixed, but the length of the other timebins is allowed to vary such that each bin forms an equal stellar mass. In the PSB model, the first three bins are fixed, the following five bins are flexible, and the most recent bin has both a variable width and variable SFR. All three models are able to accurately recover the stellar masses, metallicities, dust attenuation values, SFRs, and light-weighted ages of the mock PSBs. All three models underestimate the mass-weighted age by $\sim0.25$~dex, likely as a result of the ``outshining" problem. 

However, we see dramatic differences in how well the three models are able to recover the properties of the recent burst, particularly the quenching time. The flexible-bin model {\it always} quenches at the final bin edge, no matter what value is input by the user. In the fixed-bin model, the SFR can only change at one of the pre-chosen bin edges. Even in the most ideal case, this results in rounding errors when the quenching time falls between bins. Figure~\ref{fig:burst_props} shows that \tq is often underestimated by up to $\sim200$~Myr even when a longer \tq value is available given bin edge choices. To solve these issues, the PSB SFH model includes a final bin with {\it variable length}. This allows for the quenching timescale to be be directly informed by the data, minimizing errors due to model selection. We find that \tq values are accurately recovered by the PSB SFH model, with just $\sim90$~Myr of scatter. We confirm that the PSB SFH model provides the best fit to the data by computing the Bayes factor evidence: 99-100\% of our mock spectra show very strong preference for the PSB SFH model over any of the other three SFH models tested in this paper. 

We then test the PSB SFH model on quiescent and star-forming mock galaxies generated using the best-fit SED fitting parameters of true galaxies from the 3D-HST survey. We find that the PSB SFH model is able to recover a wide variety of ongoing SFRs. The model is also able to accurately recover the recent SFH of the galaxies, though it does overestimate the formation time (likely due to differences between our simple mock SFHs and the assumptions made by our prior). These tests indicate that the PSB SFH model is suitable for general use: it does not artificially force a large burst and a sharp quenching event.

The tests performed in this paper show that standard non-parametric models are similarly accurate at recovering basic properties of galaxies such as stellar mass, SFR and sSFR, and average age. This result suggests that standard ``out-of-the-box" non-parametric SFH models are suitable for general use, and with some caveats can be used interchangeably. However, the dramatic differences in how well the three non-parametric SFH models are able to recover \tq values indicates that small differences between these models can be critical when attempting to recover higher-order SFH quantities. Mock recovery tests such as those performed in this paper are essential to ensure that these higher-order SFH properties can be recovered accurately by a given non-parametric SFH model. We publicly provide the PSB SFH model developed in this paper as a part of \texttt{Prospector}, so that is available for the community to accurately recover the SFHs of recently-quenched galaxies.

\acknowledgements

KAS thanks Peter Behroozi, Sirio Belli, and Kevin Bundy for helpful discussions, and acknowledges the UCSC Chancellor's Postdoctoral Fellowship Program for support. {We thank the anonymous referee for a constructive, kind, and useful report.} This research used the Savio computational cluster resource provided by the Berkeley Research Computing program at the University of California, Berkeley (supported by the UC Berkeley Chancellor, Vice Chancellor for Research, and Chief Information Officer).  D.N. was funded by NSF AST-1908137.

Funding for SDSS-III has been provided by the Alfred P. Sloan Foundation, the Participating Institutions, the National Science Foundation, and the U.S. Department of Energy Office of Science. The SDSS-III web site is http://www.sdss3.org/.

SDSS-III is managed by the Astrophysical Research Consortium for the Participating Institutions of the SDSS-III Collaboration including the University of Arizona, the Brazilian Participation Group, Brookhaven National Laboratory, Carnegie Mellon University, University of Florida, the French Participation Group, the German Participation Group, Harvard University, the Instituto de Astrofisica de Canarias, the Michigan State/Notre Dame/JINA Participation Group, Johns Hopkins University, Lawrence Berkeley National Laboratory, Max Planck Institute for Astrophysics, Max Planck Institute for Extraterrestrial Physics, New Mexico State University, New York University, Ohio State University, Pennsylvania State University, University of Portsmouth, Princeton University, the Spanish Participation Group, University of Tokyo, University of Utah, Vanderbilt University, University of Virginia, University of Washington, and Yale University.

\software{astropy \citep{astropy}, scipy \citep{scipy}, seaborne \citep{seaborn}, \texttt{Prospector} \citep{johnson20}}

\bibliographystyle{aasjournal}
\bibliography{squigglebib}

\end{document}